\documentclass[usenatbib]{mn2e} 
\usepackage{amsmath}
\usepackage{graphicx}
\begin{document}

\title[Testing SPT \& ELB against simulations]{Testing standard perturbation theory and the Eulerian local biasing scheme against N-body simulations}
\author[N. Roth, C. Porciani]{Nina Roth\thanks{E-mail: nroth@astro.uni-bonn.de}, Cristiano Porciani 
\newauthor 
\footnotesize{Argelander-Institut f\"ur Astronomie der Universit\"{a}t Bonn, Auf dem H\"ugel 71, D-53121 Bonn, Germany}}

\maketitle

\begin{abstract}
We test third-order standard perturbation theory (SPT) as an approximation to non-linear cosmological structure formation. A novel approach is used to numerically calculate the three-dimensional dark matter density field using SPT from the initial conditions of two high-resolution cosmological simulations.
The calculated density field is compared to the non-linear dark matter field of the simulations both point-by-point and statistically. For smoothing scales above 8 Mpc/$h$ it shows a good agreement up to redshift 0. We present a simple fitting formula to relate the linear and non-linear density contrast that accurately recovers the non-linear time evolution for $0 \leq z \leq 10$ at the per cent level.
To address the problem of biasing between the matter field and the haloes identified in the simulation, we employ the Eulerian local bias model (ELB), including non-linear bias up to the third order. The bias parameters are obtained by fitting a scatter plot of halo and matter density (both from the simulation and from SPT). Using these bias parameters, we can reconstruct the halo density field. We find that this reconstruction is not able to capture all the details of the halo distribution. We investigate how well the large scale bias can be described by a constant and if it corresponds to the linear bias parameter $b_1$ of the local bias model. We also discuss how well the halo-halo power spectrum and the halo-mass cross spectrum from the reconstructed halo density field agree with the corresponding statistics from the simulation. 
The results show that while SPT is an excellent approximation for the matter field for suitably large smoothing scales even at redshift 0, the ELB model can only account for some of the properties of the halo density field. 
\end{abstract}

\section{Introduction}
\label{sec:intro}
Redshift surveys have revealed the existence of large-scale structures in the Universe: the galaxy distribution is organized in a complex network of filaments surrounding underdense regions and crossing at density peaks which host galaxy clusters. These structures are believed to form through gravitational instability starting from (practically) Gaussian fluctuations characterized by a nearly scale-invariant power spectrum.
Mathematically, we can follow the growth of dark-matter density (and velocity) perturbations with respect to a smooth background in terms of a set of Eulerian fluid equations coupled with the Poisson equation. As long as the density and the velocity deviate only slightly from their unperturbed values, the evolutionary equations can be linearised and solved analytically. In this case, each fluid property can be written as the superposition of a term that grows with time and a second one that decays.

At later times, however, non-linear terms in the fluid equations become important and it is no longer possible to derive an exact solution for realistic initial conditions. A widespread technique to compute approximate solutions is to use a perturbative approach and write the full solution as a series expansion in powers of the linear perturbation amplitude. This goes under the name of Standard Perturbation Theory (SPT, see \citealt{pt} for a review). In linear theory, each Fourier mode of the fluid properties evolves independently of the others. Non-linearities in the dynamics correspond to couplings between modes of different wavelengths. When the statistical properties of the initial fluctuations are known, a diagrammatic technique analogous to the Feynman diagrams can be developed to compute ensemble averaged statistics. For a given statistic (e.g. the power spectrum), the lowest-order, non-vanishing terms in the perturbative expansion give the leading expression while higher-order contributions provide additional corrections. Basically, this gives an expansion in powers of the variance of the density contrast.

Since it does not account for multi-streaming, SPT should break down in the highly non-linear configurations that lead to the formation of collapsed structures. In hierarchical scenarios for structure formation this happens at increasingly larger scales with time. An interesting question is whether, and to what degree, this breakdown affects our ability to predict the evolution of structures on the largest scales. It is generally expected that SPT is meaningful in the so called mildly non-linear regime, i.e. for Fourier modes with wavenumber $k<\ {\rm a\ few}\times 0.1\,h/ \mathrm{Mpc}$.

Comparison with N-body simulations shows that, in our currently favoured $\Lambda$CDM model and for statistics like the power spectrum and the bispectrum, SPT is rather accurate at redshifts $z>1$ (for $k<0.2 \,h/ \mathrm{Mpc}$) while it becomes increasingly imprecise as $z\to 0$ due to the fact that the variance of the density contrast approaches unity (e.g. \citealt*{2009PhRvD..80d3531C}; \citealt{2009PASJ...61..321N}). When this happens, all terms in the perturbative series are of the same importance and it does not make sense to truncate the expansion at finite order. A number of techniques have been proposed to resum or truncate the series in a more meaningful way (\citealt*{2006PhRvD..73f3519C}, \citealt*{2008JCAP...10..036P}, \citealt*{2008ApJ...674..617T}). 

Another complication arises if we want to model the large-scale structure seen in galaxy redshift surveys, namely we have to account for the fact that galaxies (and dark-matter haloes in general) are biased tracers of the underlying mass distribution. This is clearly seen observationally as different galaxy types show different clustering amplitudes. Consistently, N-body simulations show that the distribution of dark-matter haloes depends on their mass and also on other characteristics. The most common way to account for galaxy biasing is to assume that the relation between the fluctuation in galaxy counts within a characteristic volume centred at a given location can be written as a power series of the corresponding volume-averaged mass-density contrast \citep*{fry92}. This Eulerian local bias (ELB) scheme seems to give an accurate description of the large-scale clustering of dark-matter haloes in simulations \citep*{manera2009}. However, it is difficult to accurately predict the bias coefficients based on models for the collapse of fluctuations \citep*{2010MNRAS.402..589M}.

\citet*{Heavens98} have shown how the ELB model can be combined with SPT to compute the power spectrum of biased tracers of the large-scale structure up to next-to-leading order. Similarly, the dependence of higher order statistics on cosmological parameters and bias coefficients can be evaluated following the same approach (e.g. \citealt*{1994ApJ...437L..13G}; \citealt*{1997MNRAS.290..651M}; \citealt*{1999ApJ...521....1B}; \citealt*{2005PhRvD..71f3001S}).

This method has to face a technical difficulty: the ELB scheme only makes sense when density fields are averaged over a finite volume while SPT applies to unsmoothed fluctuations. Since the bias parameters depend on the actual scale used for this spatial averaging procedure the model does not make unique predictions. Also, the resulting spectra and multi-spectra show unwanted features at wavenumbers corresponding to the smoothing scale. Renormalization of the bias parameters has been proposed to alleviate the problems (\citealt{Heavens98}, \citealt{McDonald:2006mx}).

Irrespective of these difficulties, the ELB+SPT model has been used to extract the bias parameter that best fit observations (e.g. \citealt{2002MNRAS.335..432V}). Moreover, it is often used to discuss how biasing and non-linearities modify baryonic acoustic oscillations in the galaxy power spectrum (\citealt*{smith2007}; \citealt*{2006ApJ...651..619J}, 2007, 2009a) or to quantify scale-dependent biasing in the presence of non-Gaussian initial conditions (\citealt*{2008PhRvD..78l3534T}; \citealt*{2009ApJ...703.1230J}; \citealt*{2009PhRvD..80l3002S}; \citealt*{2010PhRvD..81f3530G}; \citealt*{2010arXiv1011.1513B}). Many forecasts about the constraining power of cosmological parameters from future observational campaigns are based on these calculations.

Even though each ingredient of the model has been separately tested against numerical simulations,
it still is unclear what level of accuracy can be achieved by combining SPT with the ELB scheme to describe the distribution of biased tracers of the cosmic mass distribution. Future galaxy surveys aiming at determining the origin of cosmic acceleration will require models of the galaxy power spectrum with per cent accuracy. Can the ELB+SPT computational scheme satisfy such a requirement?

In this paper, we present a direct comparison of model predictions against the outcome of state-of-the-art high-resolution N-body simulations. We follow a novel approach where we evaluate the SPT expansion of the mass density and velocity fields (up to third order) starting from the same realisation of the linear density field that has been used to generate the initial conditions of the simulations\footnote{We do not considered resummed theories for which the inclusion of biasing has only recently been considered \citep{elia2011}.}. This allows us to make a point-by-point comparison between the non-linear mass and halo overdensities while past studies have only focussed on two- or three-point statistics (of either SPT or ELB). Our analysis sheds new light on the interpretation of the bias parameters in ELB and on the effect of the smoothing procedure intrinsic to the ELB scheme.

This work is organized as follows: In section \ref{sec:SPT} we present the principles of SPT and describe how we calculate the matter density field up to third order. The numerical simulations used are described briefly in section \ref{sec:simu}. In section \ref{sec:dm} we compare the SPT matter density field to the simulations on a point-by-point basis and using one- and two-point statistics. Section \ref{sec:haloes} describes the estimation of the bias parameters, and their dependence on the halo mass and smoothing scales. The resulting halo density field is compared to the simulations, and the accuracy of the halo-halo power and halo-mass cross spectra is investigated in detail. 
We conclude in section \ref{sec:conc}.
\section{Standard Perturbation Theory}
\label{sec:SPT}
Perturbations in the matter density field can be described by the \textit{density contrast} at position $\mathbf{x}$ and conformal time $\tau$
\begin{equation}
\delta (\mathbf{x},\tau) \equiv \frac{\rho (\mathbf{x},\tau)-\bar{\rho}(\tau)}{\bar{\rho}(\tau)} ,
\label{densc}
\end{equation}
where $\bar{\rho}(\tau)$ is the mean density of the Universe, and the divergence of the \textit{velocity field} $\mathbf{v}(\mathbf{x},\tau)$
\begin{equation}
 \theta(\mathbf{x},\tau)\equiv \nabla \cdot \mathbf{v}(\mathbf{x},\tau) .
\end{equation} 
The basic idea of SPT is that the Fourier transforms, $\tilde{\delta}(\mathbf{k},\tau)$ and $\tilde{\theta}(\mathbf{k},\tau)$, can be written as a sum of separable functions of $\tau$ and $\mathbf{k}$:
\label{sec:spt}
\begin{equation}
\tilde{\delta}(\mathbf{k},\tau)= \sum_{n=1}^{\infty} D_{n}(\tau) \tilde{\delta}_{n}(\mathbf{k}),
\label{ptexp}
\end{equation}
\begin{equation}
\tilde{\theta}(\mathbf{k},\tau)= -H(\tau) \sum_{n=1}^{\infty} D_{n}(\tau) \tilde{\theta}_{n}(\mathbf{k}),
\label{ptexp2}
\end{equation}
where $D_n(\tau)\approx \left[ D_1(\tau)\right]^n$ with $D_1(\tau)$ being the linear growth factor for $\Lambda\mathrm{CDM}$ models (\citealt{bern1994}), and $D_1=1$ today.

In general, $\tilde{\delta}_{n}(\mathbf{k})$ is of $n$-th order in the linear density contrast field $\tilde{\delta}_1(\mathbf{k})$:
\begin{align}
\tilde{\delta}_{n}(\mathbf{k}) =& \int \frac{\mathrm{d}^3 q_1}{(2 \pi)^3}\  \dots \int \frac{\mathrm{d}^3 q_{n}}{(2 \pi)^3}\ \Bigl[ \tilde{\delta}_1(\mathbf{q}_1), \dots , \tilde{\delta}_1( \mathbf{q}_{n}) \nonumber \\  
&  \times F_{n}(\mathbf{q}_{\mathrm{1}}, \dots , \mathbf{q}_{n}) \  \delta_{\mathrm{D}}(\mathbf{k}-\mathbf{q_{\mathrm{1}}}- \dots -\mathbf{q_{n}}) \Bigr].
\label{ndelta}
\end{align}
The kernels $F_{n} (\mathbf{q}_{\mathrm{1}}  \dots \mathbf{q}_{n} )$ describe the mode coupling in Fourier space due to the dynamical non-linearities. They can be calculated from the recursion relations for $n \geq 2$ (first derived by \citealt{goroff1986}):
\begin{align}
F_{n} (\mathbf{q}_{\mathrm{1}},  \dots , \mathbf{q}_{n} )=&\sum_{m=1}^{n-1} \frac{G_{m}(\mathbf{q}_{\mathrm{1}},  \dots , \mathbf{q}_{m} )}{(2n+3)(n-1)} \left[ (2n+1)\alpha(\mathbf{k}_1,\mathbf{k}_2) \right. \nonumber \\
& \times F_{n-m}(\mathbf{q}_{m+1},  \dots , \mathbf{q}_{n} )  + 2\beta(\mathbf{k}_1,\mathbf{k}_2) \nonumber \\
& \left. \times G_{n-m}(\mathbf{q}_{m+1}, \dots , \mathbf{q}_{n} ) \right] \nonumber,\\
G_{n} (\mathbf{q}_{\mathrm{1}},  \dots , \mathbf{q}_{n} )=&\sum_{m=1}^{n-1} \frac{G_{m}(\mathbf{q}_{\mathrm{1}}, \dots , \mathbf{q}_{m} )}{(2n+3)(n-1)} \left[ 3\alpha(\mathbf{k}_1,\mathbf{k}_2) \right. \nonumber \\
& \times F_{n-m}(\mathbf{q}_{m+1} , \dots , \mathbf{q}_{n} ) + 2n\beta(\mathbf{k}_1,\mathbf{k}_2) \nonumber \\
& \left. \times G_{n-m}(\mathbf{q}_{m+1} , \dots , \mathbf{q}_{n} ) \right] ,
\end{align}
where $\mathbf{k}_1 \equiv \mathbf{q}_1 + \dots +\mathbf{q}_m$, $\mathbf{k}_2 \equiv \mathbf{q}_{m+1} + \dots + \mathbf{q}_n$ and $F_1 = G_1 \equiv 1$. The kernels $\alpha(\mathbf{k}_1,\mathbf{k}_2)$ and $\beta(\mathbf{k}_1,\mathbf{k}_2)$ are given by
\begin{align}
\alpha(\mathbf{k}_1,\mathbf{k}_2)& \equiv \ \frac{(\mathbf{k}_1+\mathbf{k}_2)\cdot \mathbf{k}_1}{k_1^2} \nonumber,\\
\beta(\mathbf{k}_1,\mathbf{k}_2)& \equiv \ \frac{(\mathbf{k}_1+\mathbf{k}_2)^2(\mathbf{k}_1 \cdot \mathbf{k}_2)}{2 k_1^2 k_2^2}.
\end{align}
In this work we use SPT to calculate the density contrast field up to third order. In order to meaningfully truncate the expansion after $n=3$ we need to make sure that $|\delta(\mathbf{x})| \ll 1$. This is true for very early times and/or large scales. Statistically, the SPT density field can be described by the power spectra
\begin{equation}
(2\pi)^3\,P_{mn}(k)\,\delta_{\mathrm{D}}(\mathbf{k+k'}) = \langle \tilde{\delta}_m(\mathbf{k})\ \tilde{\delta}_n(\mathbf{k'}) \rangle ,
\label{pgen}
\end{equation}
where $\langle \cdot \rangle$ denotes the ensemble average and $P_{11}(k)$ is called the \textit{linear power spectrum}.
The next-to-leading-order corrections (also called the one-loop corrections) to the power spectrum are of the order of $\tilde{\delta}_1^4$, because the correlations of third order in $\tilde{\delta}_1$ are odd moments and vanish for the case of Gaussian initial conditions. This means that the following terms contribute at the one-loop level:
\begin{equation}
P_{\textrm{1-loop}}(k) \equiv P_{\mathrm{22}}(k) + 2\,P_{\mathrm{13}}(k).
\end{equation}
The one-loop power spectra can also be directly calculated from the linear power spectrum (\citealt*{makino1992}, \citealt{Jain:1993jh}) by inserting Eq. (\ref{ndelta}) into Eq. (\ref{pgen}) and making extensive use of Wick's theorem:
\begin{equation}
P_{22}(k) = \ 2 \int \frac{\mathrm{d}^3 q}{(2 \pi)^3} F_2^2\left[\mathbf{k}-\mathbf{q},\mathbf{q} \right] P_{11}(|\mathbf{k}-\mathbf{q}|) P_{11}(q), 
\label{analyt}
\end{equation}
\begin{equation}
P_{13}(k) = \ 3\,P_{11}(k) \int \frac{d^3 q}{(2 \pi)^3}  P_{11}(q) F_3^{(s)}\left[\mathbf{k}, \mathbf{q}, -\mathbf{q} \right],
\label{analyt2}
\end{equation}
where $F_3^{(s)}\left[\mathbf{k}, \mathbf{q}, -\mathbf{q} \right]$ denotes $F_3\left[\mathbf{k}, \mathbf{q}, -\mathbf{q} \right]$ symmetrized w.r.t. its arguments. 

In previous studies, the one-loop corrections to the power spectrum were calculated using Eqs. (\ref{analyt}) and (\ref{analyt2}). In this work we explicitly calculate the third-order density contrast field \textit{point-by-point} on a 3D grid and compare it to a simulated non-linear density contrast field to test the validity of SPT at low redshifts. We start from an initial density field with $N^3$ grid points and calculate the corresponding higher order density fields, which can then be evolved to any redshift according to Eq. (\ref{ptexp}).

While Eq. (\ref{ndelta}) is an elegant analytical description of how the higher order density contrast field depends on the linear one, there are other formulations which are more suited for numerical integrations, because they do not require $(n-1)$ integrations over $k$-space (the integral over the Dirac-delta-function can be done analytically). \cite{makino1992} showed that Eq. (\ref{ndelta}) can be rewritten as a recursion relation (their Eqs. 2.15a and 2.15b), expressed schematically like:
\begin{align}
\tilde{\delta}_n & =K\left[\tilde{\delta}_1,\dots,\tilde{\delta}_{n-1}, \tilde{\theta_1},\dots, \tilde{\theta}_{n-1}\right]\nonumber,\\ 
\tilde{\theta}_n & =J\left[\tilde{\delta}_1,\dots,\tilde{\delta}_{n-1}, \tilde{\theta}_1,\dots,\tilde{\theta}_{n-1}\right],
\end{align}
where $K$ and $J$ are \textit{single} integrals over $k$-space with (different) mode-coupling kernels. The starting point of this integration is given by the linearised continuity equation: $\tilde{\theta}_1(\mathbf{k})=\tilde{\delta}_1(\mathbf{k})$. This method is numerically faster because it takes significantly less time to calculate two single integrals (e.g. $\tilde{\delta}_2$ and $\tilde{\theta}_2$ which are needed for $\tilde{\delta}_3$) than to perform a double-integral for $\tilde{\delta}_3$ using Eq. (\ref{ndelta}). This also means that even higher-order corrections could be calculated in the same amount of time and one obtains the velocity divergence field $\tilde{\theta}(\mathbf{k})$ up to the $(n-1)$-th order as a side product. We find that this method is also more stable against numerical effects such as the discretization of the mode-coupling kernels.
\begin{figure*}
\centering
\includegraphics[scale=0.35]{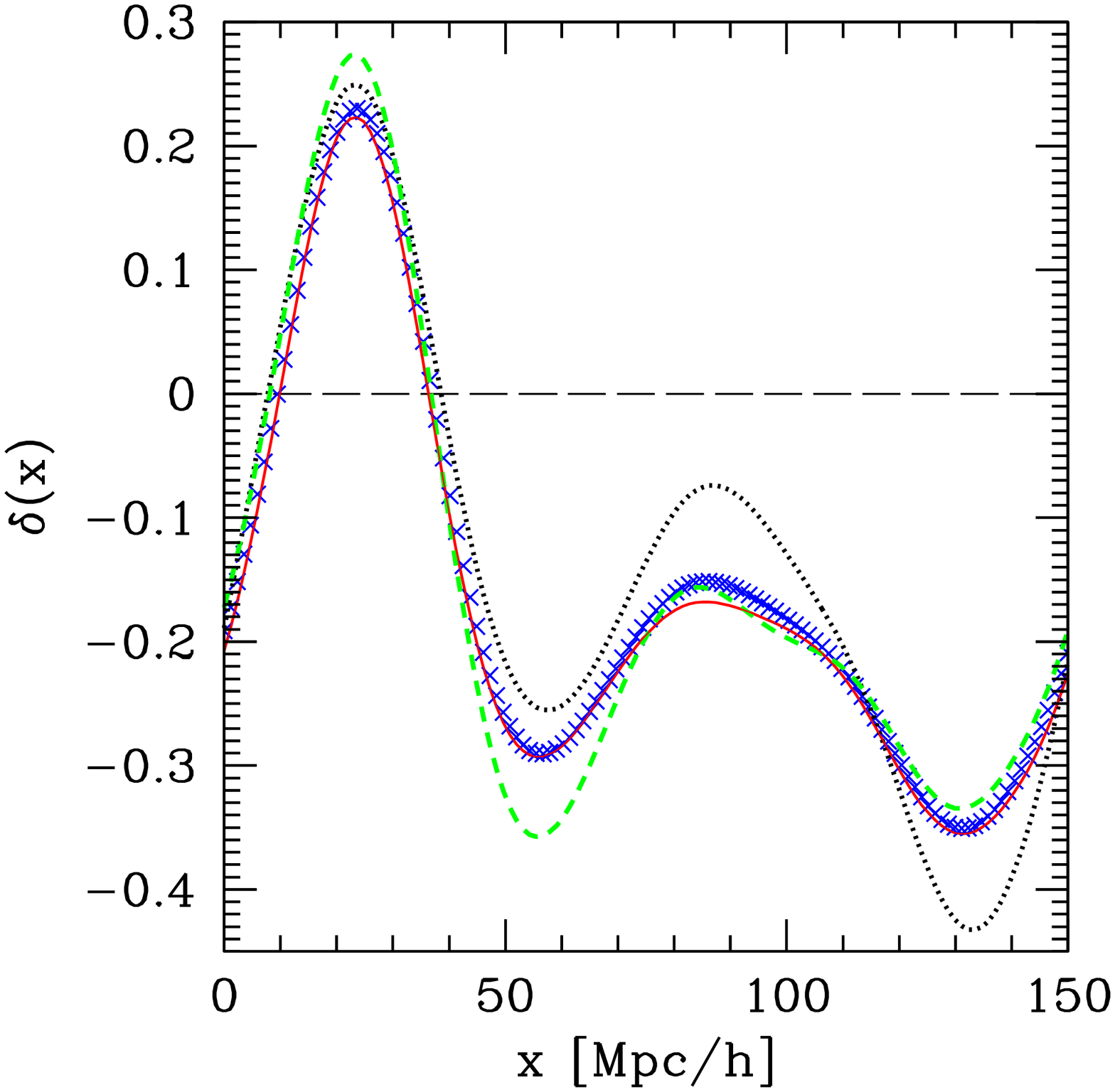}
\includegraphics[scale=0.35]{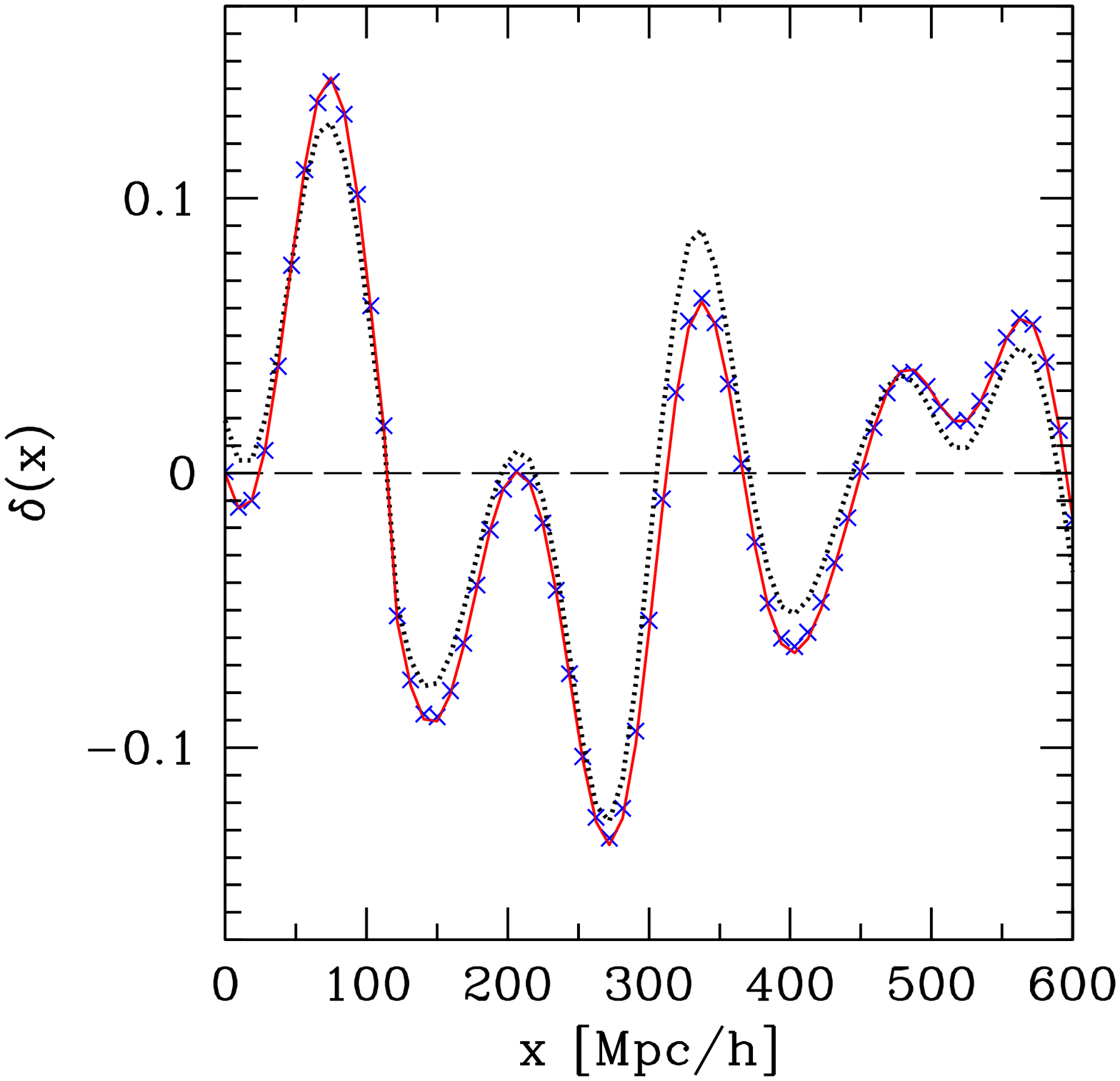}
\caption{Left panel: $\delta_{\mathrm{m}}^{\mathrm{s}}(\mathbf{x})$ (blue crosses) compared to the different SPT orders (black dotted: $\delta_1^{\mathrm{s}}(\mathbf{x})$, green dashed: $\delta_1^{\mathrm{s}}(\mathbf{x})+\delta_2^{\mathrm{s}}(\mathbf{x})$, red solid: $\delta_{\mathrm{SPT}}^{\mathrm{s}}(\mathbf{x})$) at redshift 0 for the small box with $R=12\ \mathrm{Mpc}/h$. Right panel: Same for the large box and $R=28\ \mathrm{Mpc}/h$. (Second order is not shown here because it is very similar to the third order for this smoothing scale.)}
\label{fig:dxptvsdxm_high}
\end{figure*}

However, even with this hierarchical method, the integration time still scales with the square of the number of grid points, because if we increase the number of grid points by, say, a factor of 8 (doubling the resolution in each dimension), for each of these points we also have 8 additional grid points that contribute to each integral. This means we are limited by the resolution of our grid, i.e. the minimum separation between grid points. The highest resolution that can be achieved with a common workstation is $(128)^3$ grid points.
\section{N-body Simulation}
\label{sec:simu}
The simulations that are used in this work are described in detail in \citet*{Pillepich08}. They were run using the TreePM code \textit{Gadget-2} (\citealt{Springel2005}) in a flat $\Lambda\mathrm{CDM}$ cosmology starting from a Gaussian random field. The cosmological parameters are $\Omega_{\mathrm{m}}=0.279$, $\Omega_{\mathrm{b}}=0.0462 $, $\Omega_{\Lambda}=0.721$, $h= 0.701$, $\sigma_8=0.817$ and  $n_{\mathrm{s}}=0.96$. The output contains the 3D positions and velocities of the dark matter particles, which are converted into a density contrast field on a grid using the cloud-in-cell (CIC) algorithm. Outputs for 30 time steps between $z=10$ and $z=0$ logarithmically spaced in $(1+z)^{-1}$ are available. The linear density contrast field $\delta_1 (\mathbf{x})$, which is the basis of our calculations, is obtained from the initial conditions of the simulation. The non-linear matter field from the simulation output at $z=0$ we denote as $\delta_{\mathrm{m}}(\mathbf{x})$.

The simulations consist of two cubic volumes with different side lengths $L$, which are used to explore different length scales and halo masses. The small box has $L=150\ \mathrm{Mpc}/h$ and contains $(1024)^3$ dark matter particles with mass $M_{\mathrm{part}}=2.433 \times 10^8\, \mathrm{M}_{\odot}/h$. The large box has $L=1200\ \mathrm{Mpc}/h$ and $(1024)^3$ particles with $M_{\mathrm{part}}=1.246 \times 10^{11}\, \mathrm{M}_{\odot}/h$. There are a total of 1,051,230 (1,953,437) haloes at redshift 0 identified with the Friends-of-Friends algorithm in the small (large) simulation volume. The linking length is 0.2 times the mean inter-particle separation and each halo contains at least 100 particles. This gives a total mass range of $2.433 \times 10^{10} M_{\odot}/h \leq M_{\mathrm{h}} \leq 1.2 \times 10^{15} M_{\odot}/h$. The initial redshifts are $z=70$ and $z=50$, respectively.

To compare the SPT results with the simulations, we need to smooth the density field. We adopt a Gaussian kernel, which corresponds to a simple multiplication in Fourier space, and denote the smoothed fields with a superscript s:
\begin{equation}
\tilde{\delta}^{\mathrm{s}}(\mathbf{k})
\equiv \mathrm{e}^{\frac{-(|\mathbf{k}| R)^2}{2}} \tilde{\delta}(\mathbf{k}).
\label{smooth}
\end{equation}
The resolution limit of $(128)^3$ grid points for the SPT calculation restricts the scales that can be probed for the large simulation box, because each grid cell has a minimal size of $1200/128\ \mathrm{Mpc}/h \approx 9.4\ \mathrm{Mpc}/h$. In order for the smoothing kernel to extend over several grid cells, we have to choose $R\approx 28\ \mathrm{Mpc}/h$ for the large simulation volume. For the small box, the resolution restriction is not significant, but we find that SPT does not give accurate results if $\delta^{\mathrm{s}}(\mathbf{x})$ is of order unity at redshift 0. For this reason, we adopt $R=12\ \mathrm{Mpc}/h$ which corresponds to an r.m.s. density contrast of 0.6.

We will use the output of the simulations for both dark matter and dark matter haloes to compare with our SPT calculation and investigate different length scales, halo masses and redshifts. The simulation results are of course not limited by the SPT grid resolution, but if we want to make point-by-point comparisons, we have to use the same grid and smoothing scale to the simulated mass and halo distributions.

\section{Dark Matter} 
\label{sec:dm}
\begin{figure*}
\centering
\includegraphics[scale=0.35]{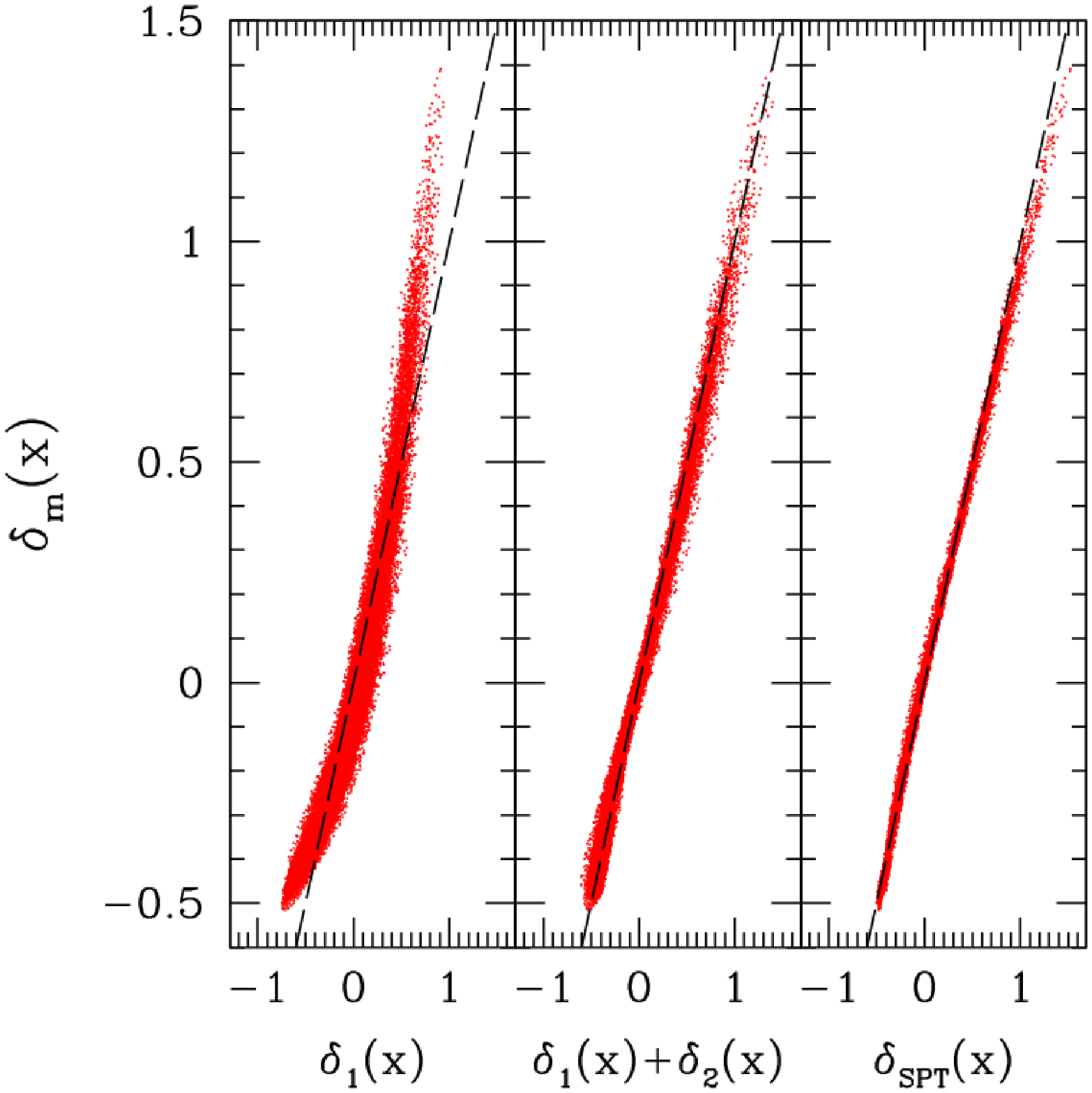}
\includegraphics[scale=0.35]{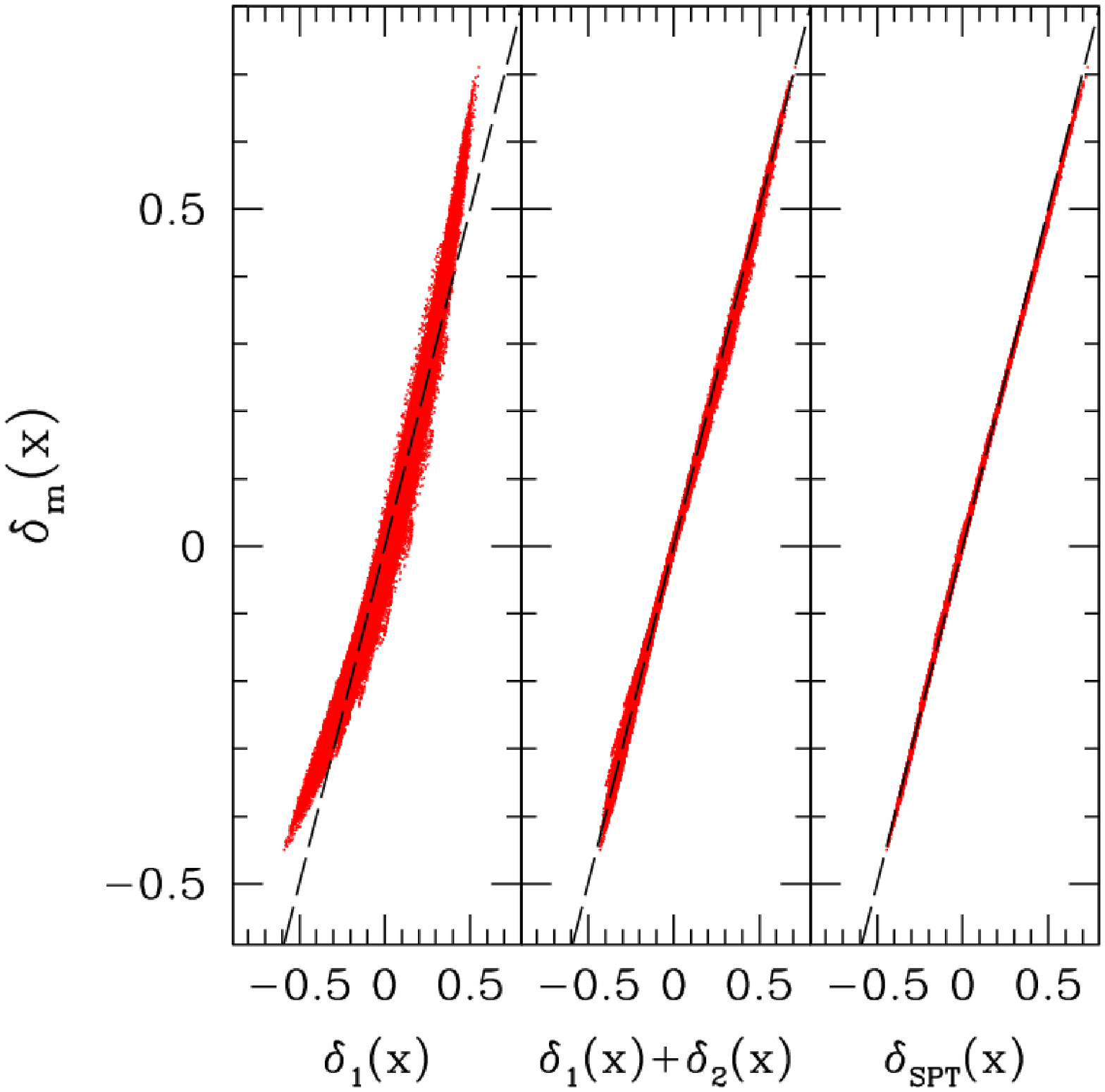}
\caption{Left: Scatter plot of $\delta_{\mathrm{m}}^{\mathrm{s}}(\mathbf{x})$ vs. linear order SPT (left panel), second-order SPT (middle panel), third-order SPT (right panel). Black dashed lines show $\delta_{\mathrm{m}}^{\mathrm{s}}/\delta_{\mathrm{SPT}}^{\mathrm{s}}=1$. The smoothing scale is $R=12\ \mathrm{Mpc}/h$. Note that we only show one out of eight points to improve readability. Right: Same for the large box with $R=28\ \mathrm{Mpc}/h$.}
\label{fig:dptvsdm}
\end{figure*}
In the following sections, all density fields are considered to be at redshift 0, unless specifically stated otherwise.
\subsection{SPT Density Contrast Field}
\label{sec:sptdenscon}
We want to approximate the non-linear matter density contrast field $\delta_{\mathrm{m}}^{\mathrm{s}}(\mathbf{x})$ with the third-order SPT field $\delta_{\mathrm{SPT}}^{\mathrm{s}}(\mathbf{x})\equiv \delta_1^{\mathrm{s}}(\mathbf{x})+\delta_2^{\mathrm{s}}(\mathbf{x})+ \delta_3^{\mathrm{s}}(\mathbf{x})$. We also have to investigate if the third-order expansion is actually more accurate than the lower orders. Fig. \ref{fig:dxptvsdxm_high} shows the different approximations to $\delta_{\mathrm{m}}^{\mathrm{s}}(\mathbf{x})$ along a line parallel to one of the coordinate axis in the simulation over the box-length $L$ (half a box-length for $L=1200\ \mathrm{Mpc}/h$ to make the lines more distinguishable). For both simulation volumes, the third-order density contrast gives the best approximation to the simulated matter density contrast $\delta_{\mathrm{m}}^{\mathrm{s}}(\mathbf{x})$. The agreement is already pretty good for the small box but excellent for the large box, owing to the larger smoothing scale in the latter case. While the linear density contrast $\delta_1^{\mathrm{s}}(\mathbf{x})$ traces the overall structure of the non-linear field, the higher order expansion is more accurate.

Fig. \ref{fig:dptvsdm} shows a scatter plot of the non-linear field in the simulation and the different SPT orders for both volumes. Starting from the left panel, we see again that the linear density contrast is not a very good approximation to the non-linear density contrast. Although this result is not really surprising, the plot once again shows that the linear approximation is especially wrong when $|\delta^{\mathrm{s}}(\mathbf{x})| \not\ll 1$. The middle panel shows the density contrast up to second order and the right panel shows the density contrast up to third order. Note that the third-order over corrects the linear density around $\delta^{\mathrm{s}} \approx 1$. From left to right the scatter around the black line given by $\delta_{\mathrm{m}}^{\mathrm{s}}/\delta_{\mathrm{SPT}}^{\mathrm{s}}=1$ is reduced. In all three panels the difference to $\delta_{\mathrm{m}}^{\mathrm{s}}(\mathbf{x})$ is largest when $|\delta^{\mathrm{s}}(\mathbf{x})| \not\ll 1$. Again, the difference in scatter between the large and small simulation volumes is due to the different smoothing scales used (12 and 28 Mpc$/h$). 

\begin{figure}
\centering
\includegraphics[scale=0.35]{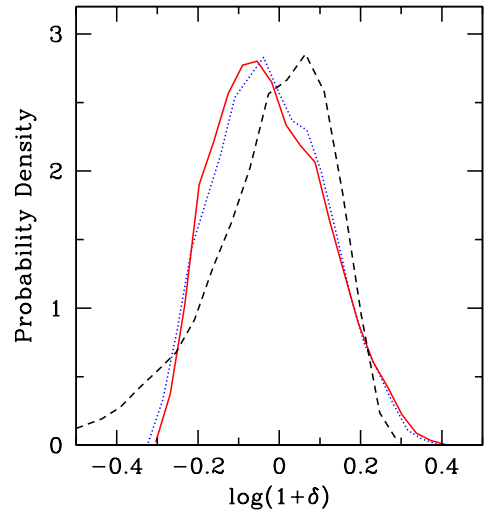}
\caption{Probability distribution function (PDF) for $\delta_{\mathrm{SPT}}^{\mathrm{s}}(\mathbf{x})$ (red solid), $\delta_{\mathrm{m}}^{\mathrm{s}}(\mathbf{x})$ (blue dotted) and $\delta_1^{\mathrm{s}}(\mathbf{x})$ (black dashed) at redshift 0 (small box, $R=12\ \mathrm{Mpc}/h$).}
\label{fig:pdf_matter}
\end{figure}

Fig. \ref{fig:pdf_matter} shows the probability distribution function (PDF) for the linear, SPT and non-linear matter density contrast for the small box, with a smoothing scale $R=12\ \mathrm{Mpc}/h$. This also shows that SPT provides a good approximation for the non-linear density contrast even at redshift 0.
\subsubsection{Redshift Dependence}
\label{ssec:time}
\begin{figure*}
\centering
\includegraphics[scale=0.35]{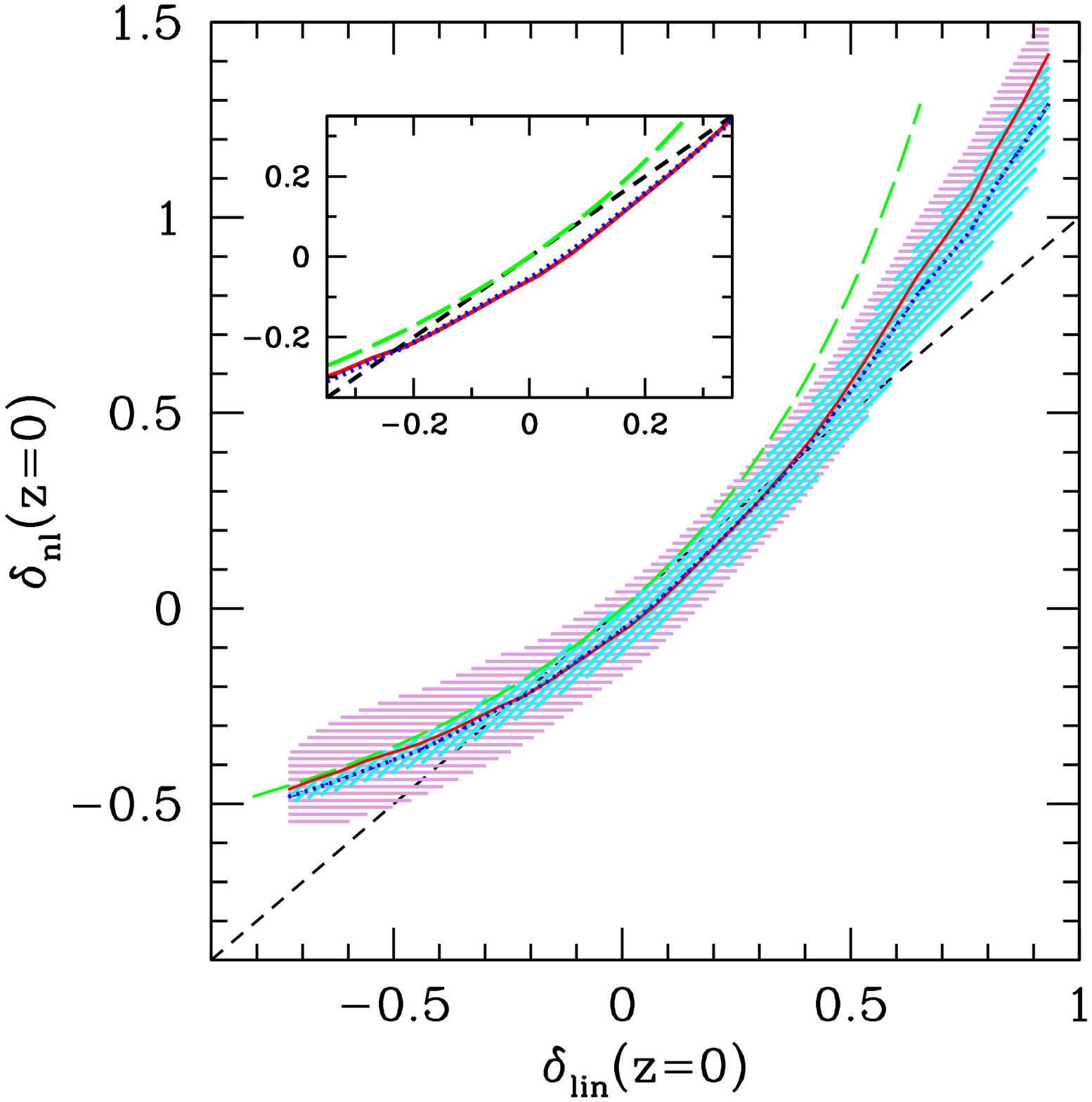}
\includegraphics[scale=0.35]{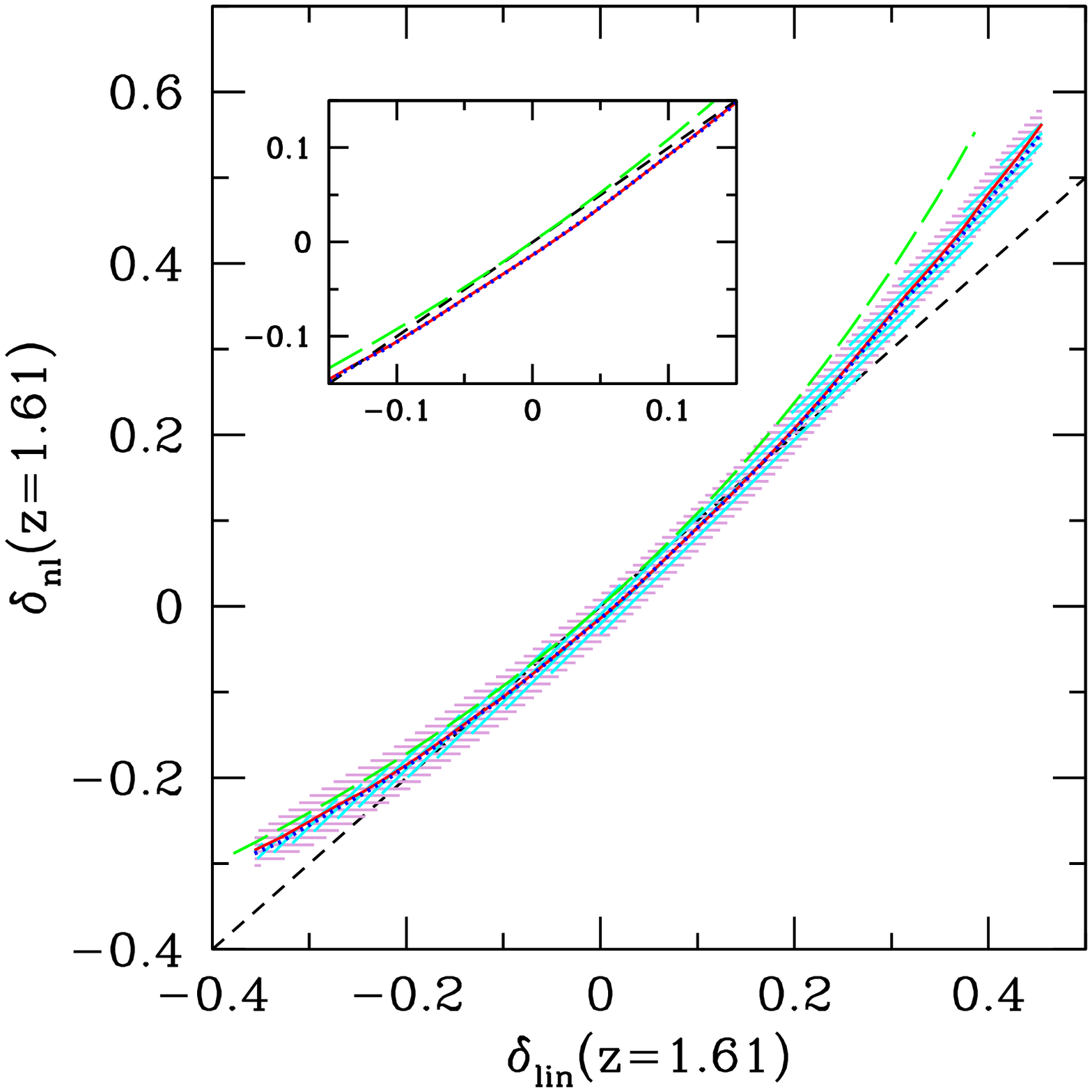}
\caption{Left: Mean non-linear matter density contrast from the simulation (blue dotted), SPT (red solid) and an analytic prediction from the spherical collapse model (green long-dashed) as a function of the linear density contrast at redshift 0, for the small box with $R=12\ \mathrm{Mpc}/h$. The hatched areas show the 1-$\sigma$ scatter around the mean for  $\delta_{\mathrm{SPT}}^{\mathrm{s}}$ (horizontal hatching) and $\delta_{\mathrm{m}}^{\mathrm{s}}$ ($45^{\circ}$), and the black short-dashed line corresponds to linear evolution. The inset shows the region around $\delta^{\mathrm{s}}=0$, where linear theory actually overpredicts the non-linear evolution. Right: Same at $z=1.61$.}
\label{fig:time2}
\end{figure*}
So far we have only compared our SPT calculation at redshift 0. Now we want to extend the comparison to higher redshifts. Fig. \ref{fig:time2} shows the conditional mean of the non-linear density contrast given its linear counterpart at redshift 0 (left panel) and redshift 1.6 (right panel). The lines show the conditional mean density corresponding to: the linear evolution (black short-dashed), the SPT density (red solid), the non-linear evolution from the simulation (blue dotted) and an analytic expectation from the spherical collapse model (\citealt*{1994A&A...291..697B}, \citealt*{1996MNRAS.282..347M}, green long-dashed), all smoothed with $R=12\ \mathrm{Mpc}/h$. The hatched areas show the 1-$\sigma$ scatter contours for SPT and simulation. SPT and non-linear matter density agree well if $\delta_{\mathrm{lin}}^{\mathrm{s}}$ is not too large, as expected, but the scatter around the mean SPT density is generally larger than around $\delta_{\mathrm{m}}^{\mathrm{s}}$. The spherical collapse model (which was derived for the Einstein-de Sitter cosmological model) always overpredicts the non-linear density contrast, and gets close only when $\delta^{\mathrm{s}}<0$. The linear approximation overpredicts the simulation at low densities, and underpredicts it at high densities. The right panel shows the same for redshift 1.61. At this redshift, SPT and non-linear matter density are in almost perfect agreement, but the scatter in SPT is still slightly larger. The spherical collapse model and the linear evolution are now closer to the non-linear evolution, but still show the systematics seen at redshift 0.

The spherical collapse model is commonly used to compute Eulerian bias coefficients for dark matter haloes (e.g. \citealt{1996MNRAS.282..347M}, \citealt{2010PhRvD..81f3530G} and references therein), but it does not describe our data well. We therefore provide a new and more accurate fitting formula for the relation between $\delta_{\mathrm{m}}^{\mathrm{s}}$ and $\delta_{\mathrm{1}}^{\mathrm{s}}$. The redshift range we can study is $0 \leq z \leq 10$ in 30 bins, logarithmically spaced in $(1+z)^{-1}$. For each redshift bin, we find that the parameterisation
\begin{equation}
\delta_{\mathrm{m}}(z) = A(z) + B(z) \cdot \delta_{\mathrm{1}}^{\mathrm{s}}(z) + C(z) \cdot [\delta_{\mathrm{1}}^{\mathrm{s}}(z)]^2
\label{fit_nl_pol}
\end{equation}
gives a very good fit for all values of $\delta_1^{\mathrm{s}}$: the ratio of the fit residuals to the true value is less than 5 per cent for redshift 0 and less than 1 per cent for $z>6$. The functions $A$, $B$ and $C$ depend on $z$ in the following way:
\begin{align}
A(z)&= 0.006 - \frac{0.054}{1+z}\nonumber, \\
B(z)&= 1.000 - \frac{0.045}{1+z}\nonumber ,\\
C(z)&= 0.643 - 0.012\, z - \frac{0.122}{1+z}.
\label{fit_func_xdep}
\end{align}
In the limiting case $z \rightarrow \infty$, some of the fit coefficients do not show the right asymptotic behaviour ($A \rightarrow 0$, $B \rightarrow 1$, $C \rightarrow 0$) as one would expect because $\delta_{\mathrm{m}}^{\mathrm{s}} \rightarrow \delta_{\mathrm{1}}^{\mathrm{s}}$, so this fit should \textit{not} be used for redshifts outside the fitted range. Both density fields were smoothed with $R=12\ \mathrm{Mpc}/h$. The choice of $R$ does not significantly influence the fitting parameters, but the agreement is worse when large values of $\delta_1^{\mathrm{s}}$ and $\delta_{\mathrm{m}}^{\mathrm{s}}$ are allowed (i.e. $R$ is small at low redshifts).

Another relation between the linear and non-linear density fields that is sometimes used is the lognormal transformation (e.g. \citealt*{1991MNRAS.248....1C}, \citealt*{1995ApJ...443..479B}). This has proven to yield a PDF which agrees well with simulations. However, this does not imply that it can be used as a point-by-point relation between the linear and non-linear density fields, as pointed out by \citet*{2001ApJ...561...22K}. We follow their parameterisation, which can be rewritten as
\begin{equation}
\delta_{\mathrm{m}}^{\mathrm{s}}(z) = e^{\gamma(\delta_1^{\mathrm{s}}(z)- \frac{1}{2}\gamma \sigma^2)}-1,
\end{equation}
with
\begin{equation}
\gamma^2 \sigma^2 = \ln ( 1+\sigma^2_{\mathrm{m}} ).
\end{equation}
Here, $\sigma^2$ ($\sigma^2_{\mathrm{m}}$) is the variance of the smoothed linear (non-linear) density field. We calculate $\sigma^2$ and leave $\gamma$ as the fit parameter. This fit yields values for $\delta_{\mathrm{m}}$ which are systematically lower by $\approx 30$ per cent at redshift 0 and still 20 per cent at $z>6$. So indeed the lognormal transformation can not be used to accurately predict the evolution of an initial Gaussian field into a non-linear field. Again, this does not mean that the statistics of a non-linear density field can not be described by a lognormal field, just that one can not expect that, for a point-by-point comparison, the linear field used in the transformation corresponds to the initial conditions of that non-linear field.
\subsection{Matter Power Spectra}
\label{ssec:mps}
\begin{figure}
 \centering
\includegraphics[scale=0.39]{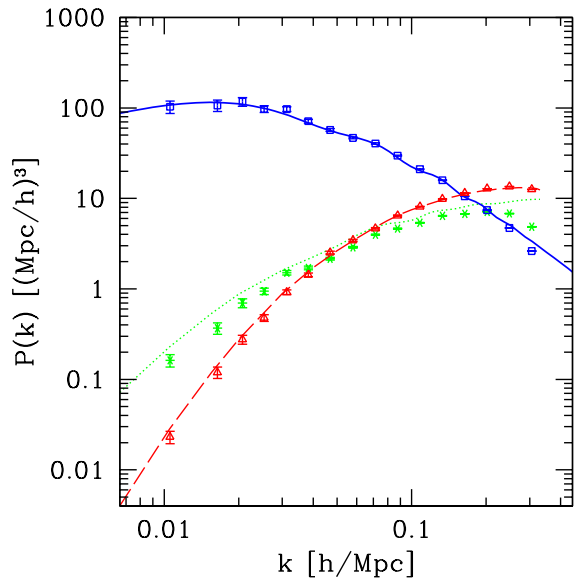}
 \caption{Contributions to the SPT matter power spectrum for the big box: $P_{11}(k)$ (blue solid line and squares), $P_{22}(k)$ (red dashed line and triangles) and $|2\,P_{13}(k)|$ (green dotted line and stars). The points are the results from our calculations, while the lines are obtained from (integrating) a linear power spectrum (according to Eq. \ref{analyt}).}
 \label{fig:p11theo}
\end{figure}

After calculating the higher order density contrast, we now calculate the matter power spectra up to the 1-loop order. We can do this in two ways: calculating the volume average of the SPT density contrast (Eq. \ref{pgen}) or integrating a given linear power spectrum $P_{11}(k)$ (Eq. \ref{analyt}), where $P_{11}(k)$ is the same power spectrum that was used to set up the initial conditions of the simulation. This also serves as a consistency check for our grid-based calculation. The results for the large box are shown in Fig. \ref{fig:p11theo}. The small box is not shown because it is very similar. The lines show the results from Eq. (\ref{analyt}) and the points show the numerical result, without smoothing, i.e. $R=0$. The errors of the power spectra are calculated assuming that our fields are Gaussian:
\begin{equation}
\sigma^2_P = \frac{2}{n_{\mathrm{B}}} P^2(k),
\end{equation}
where $n_{\mathrm{B}}$ is the number of modes in each $k$-bin.

The linear power spectrum from the simulation agrees very well with the input power spectrum until the scales which are affected by the resolution of the grid of the matter field. Once we apply smoothing, this small-scale effect will disappear. The $P_{22}$-term from the integration and the SPT calculation fits the simulation very well on all applicable scales. Finally, the $P_{13}$-term from the SPT calculation is slightly lower than the power spectrum integration (a factor of $\approx 1.1$). The lines that are shown in Fig. \ref{fig:p11theo} are the integration result with limits corresponding to the scales which are available in our simulation box, but since the box only contains discrete modes and the power spectrum integration assumes a continuous frequency spectrum, choosing the same upper and lower bound for both methods does not necessarily give the same result. A consistency check for both methods is if the power spectra overlap in the $k$-region covered by both boxes. We find that the grid-based method gives very consistent results in this region, while there is an offset for the power spectrum integration. By slightly modifying the integration limits for $P_{13}(k)$, the discrepancy in the latter can be fixed, and both methods agree with each other.

\begin{figure}
 \centering
 \includegraphics[scale=0.39]{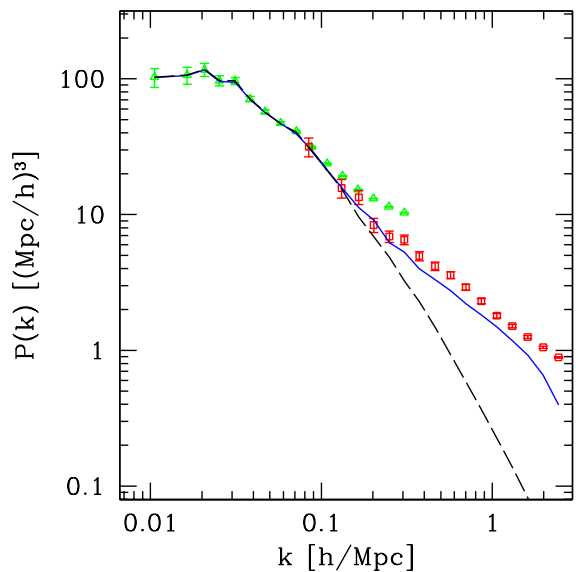}
 \caption{The third-order SPT power spectrum (points with error bars) compared to the non-linear power spectrum from the simulation (blue solid line) and the linear power spectrum evolved to redshift 0 (black dashed line) for both boxes. Green triangles and red squares are used to distinguish the large and small box.}
 \label{fig:psum}
\end{figure}

Fig. \ref{fig:psum} shows the full third-order SPT matter power spectrum compared to the simulation. The lines show the power spectra directly calculated from the linear and non-linear density contrast in the simulation, which have been ``glued together" to show both volumes at the same time. The points show the result from our grid-based calculation. For the large box, linear theory, SPT and non-linear evolution agree until $k \approx 0.1\, h/\mathrm{Mpc}$, but linear theory seems to be the better approximation for slightly larger $k$. For the small box, the SPT matter power spectrum describes the non-linear evolution better than the linear power spectrum for scales $0.2\, h/\mathrm{Mpc}<k< 0.3\, h/\mathrm{Mpc}$, while they are equivalent for larger scales.
We stress again that these power spectra (or the density contrast used to compute them) are not smoothed at all in Fig. \ref{fig:p11theo} and Fig. \ref{fig:psum}. This explains the discrepancy of the SPT power spectra on small scales with respect to the good agreement of the density contrasts shown in Fig. \ref{fig:dxptvsdxm_high}. The fact that the points deviate from each other in the overlap region between the large and small box is due to $P_{13}(k)$, as discussed above. There is excess power on these scales for the full SPT power spectrum because $P_{13}(k)< 0$.
\section{Haloes}
\label{sec:haloes}
In this section we study dark matter haloes and how they relate to the underlying matter density field. The haloes are split up in 3 mass bins for each box because the bias parameters are very sensitive to the mass of the objects (see also Fig. \ref{fig:massdep}). Table \ref{tab:bins} shows the selection criteria for the binning. Halo positions are assigned to the grid using the CIC algorithm to obtain a continuous halo density contrast field $\delta_{\mathrm{h}}(\mathbf{x})$.
\begin{table}
\centering
\caption{Halo mass bins: I-III for the small box and IV-VI for the large box.}
\begin{tabular}{|c|c|c|}
\hline Bin & Mass range [$10^{12} M_{\odot}/h$] & \# of haloes \\ 
\hline 
I & 0.025 - 0.035 & 91,511 \\ 
II & 0.5 - 2  & 17,030 \\ 
III & 5 - 10 & 2,693 \\
IV & 12.5 - 30 & 376,292\\ 
 V & 60 - 150 & 71,581\\
VI & 300 - 1200 & 7,673\\ 
\hline 
\end{tabular}
\label{tab:bins}
\end{table}
The bins were chosen such that they are separated in mass, but still contain a large enough number of haloes to reduce the effect of shot noise.
\subsection{Bias Estimation}
\label{ssec:bias}
Many different bias models exist in the literature. Here, we use the \textit{local Eulerian} bias model
\begin{equation}
\delta_{\mathrm{h}}(\mathbf{x})=F\left[\,\delta_{\mathrm{m}}(\mathbf{x}) \,\right] ,
\label{b1}
\end{equation}
implying that the halo density $\delta_{\mathrm{h}}(\mathbf{x})$ depends only on the matter density $\delta_{\mathrm{m}}(\mathbf{x})$ at the same point in space through a general function $F$. A choice of $F$ which is widely used is that of a linear relation between $\delta_{\mathrm{h}}(\mathbf{x})$ and $\delta_{\mathrm{m}}(\mathbf{x})$ with a bias factor $b$:
\begin{equation}
\delta_{\mathrm{h}}(\mathbf{x}) = b \, \delta_{\mathrm{m}}(\mathbf{x}).
\end{equation}
This so-called linear biasing only changes the amplitude but not the shape of the power spectrum, since $P_{\mathrm{h}}(\mathbf{k}) = b^2 \, P_{\mathrm{m}}(\mathbf{k})$. It has been shown that this is not in agreement with the power spectra derived from numerical simulations (e.g. \citealt{smith2003}). Also, for $|b| > 1$, it can lead to $\delta_{\mathrm{h}}(\mathbf{x}) < -1$, which of course does not make physical sense. However, on \textit{large scales} the linear approximation is valid as will also be shown in section \ref{ssec:beff}.

The next step would be to consider a \textit{non-linear} relation between $\delta_{\mathrm{h}}(\mathbf{x})$ and $\delta_{\mathrm{m}}(\mathbf{x})$. If we assume that $\delta_{\mathrm{m}}(\mathbf{x})$ is a smooth field, we can Taylor-expand the right hand side of Eq. (\ref{b1}) around $\delta_{\mathrm{m}}(\mathbf{x}) \approx 0$ and get an expression for the higher-order \textit{bias coefficients} $b_j$
\begin{equation}
\delta_{\mathrm{h}}(\mathbf{x})=\sum_{j=0}^{\infty} \frac{b_{j}}{j!}\, \delta^{j}_{\mathrm{m}}(\mathbf{x}) ,
\label{biasex}
\end{equation}
where $b_1$ is the linear bias, and $b_0$ is determined by the requirement $\langle \delta_{\mathrm{h}} (\mathbf{x}) \rangle=0$. To be consistent with our third-order SPT matter field, we first smooth $\delta_{\mathrm{h}}(\mathbf{x})$ according to Eq. (\ref{smooth}) and truncate Eq. (\ref{biasex}) after $b_{\mathrm{3}}$:
 \begin{equation}
 \delta_{\mathrm{h}}^{\mathrm{s}}(\mathbf{x}) 
= \ b_1 \delta_{\mathrm{m}}^{\mathrm{s}}(\mathbf{x}) +\frac{b_2}{2} \left( \delta_{\mathrm{m}}^{\mathrm{s}}(\mathbf{x}) \right)^2  + \frac{b_3}{6} \left( \delta_{\mathrm{m}}^{\mathrm{s}}(\mathbf{x}) \right)^3.
\end{equation}
This means we obtain the $b_i$ by fitting a polynomial to the scatter plot of $\delta_{\mathrm{h}}^{\mathrm{s}}(\mathbf{x})$ and $\delta_{\mathrm{m}}^{\mathrm{s}}(\mathbf{x})$.

However, there is a subtlety in the way the local bias model is applied. While the input mass density contrast field is smoothed on a certain scale $R$, and also the halo density field in the scatter plot is smoothed on the same scale, the \textit{reconstructed} halo density from the fits also contains contributions from smaller scales. The reason for this is the choice of the filter function in Eq. (\ref{smooth}). If we used a top-hat filter in $k$-space, this effect would vanish, because then all modes with $k>R$ are set to zero (however this filter oscillates in real-space). But with the Gaussian filter, modes with, say, $kR=2$ are only damped by a factor of $\approx 0.14$. When applying the ELB model, taking the square and cube of the smoothed field in real-space corresponds to a convolution in $k$-space. This \textit{unsmoothing effect} can be seen especially in the halo power spectra (section \ref{sec:halops}), and limits the scales where we can trust the calculated halo power spectra even more than the validity of the SPT approximation discussed in section \ref{ssec:mps}.

\begin{figure*}
\centering
\includegraphics[scale=0.35]{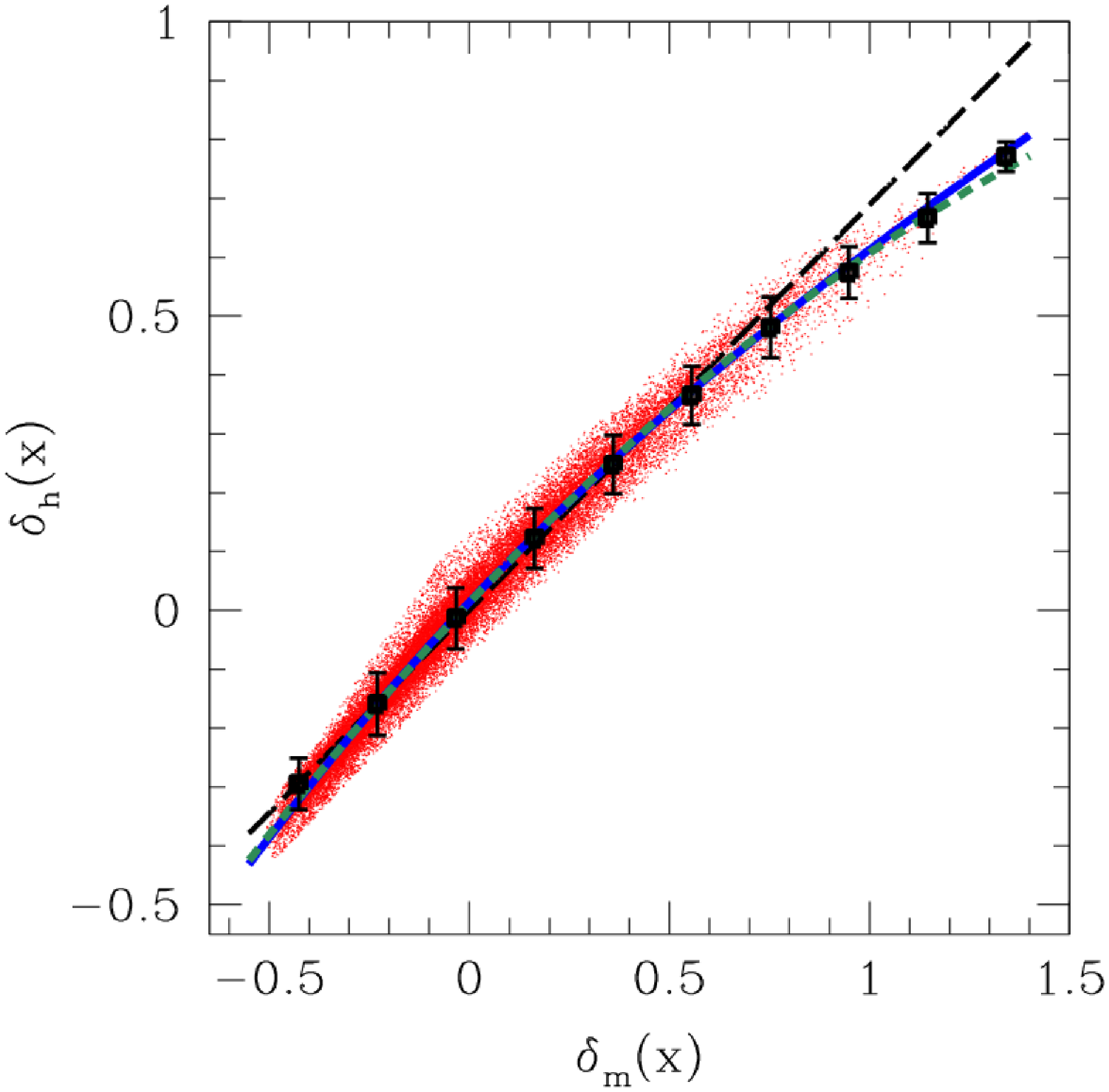}
\includegraphics[scale=0.35]{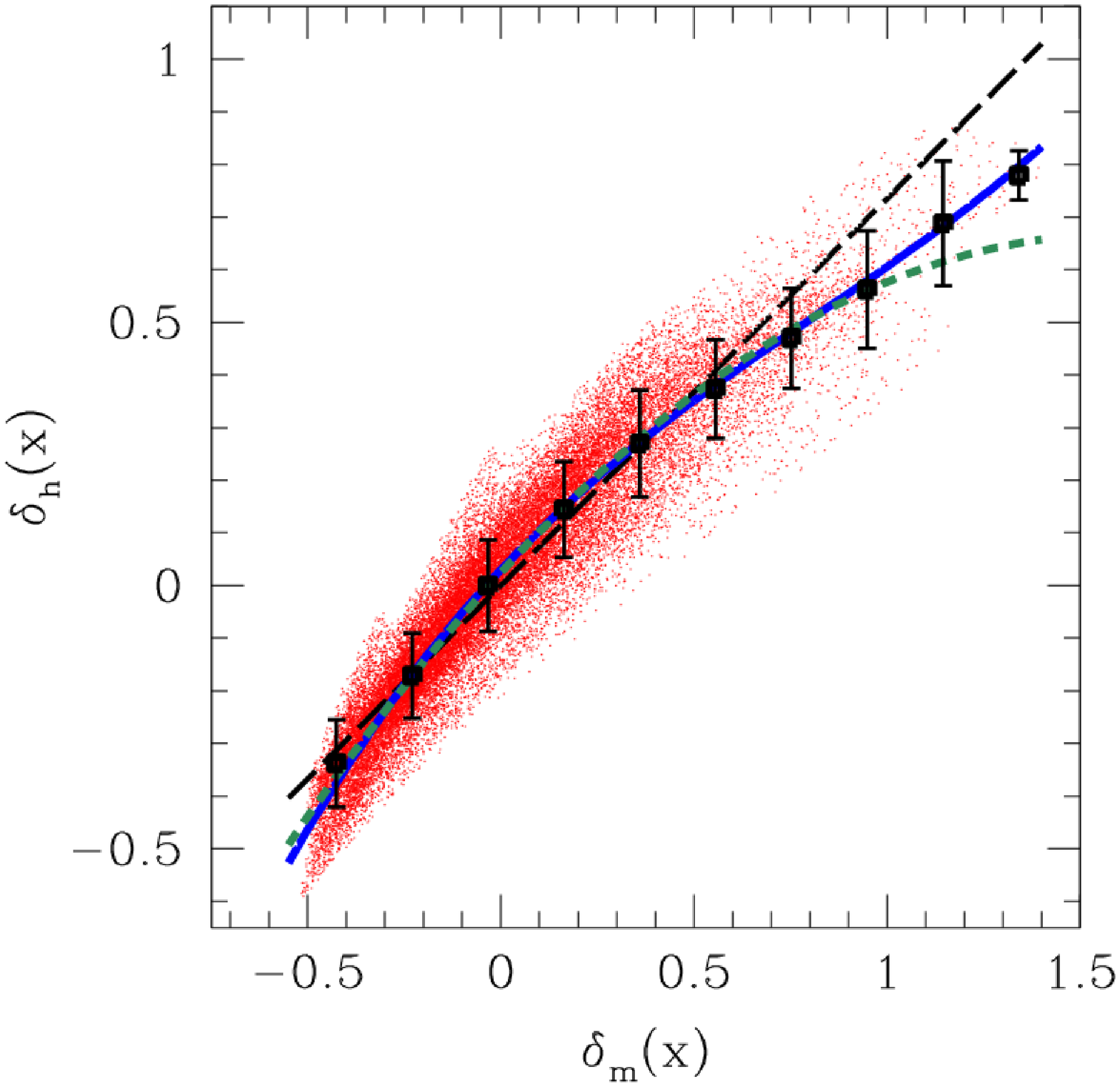}\\
\includegraphics[scale=0.35]{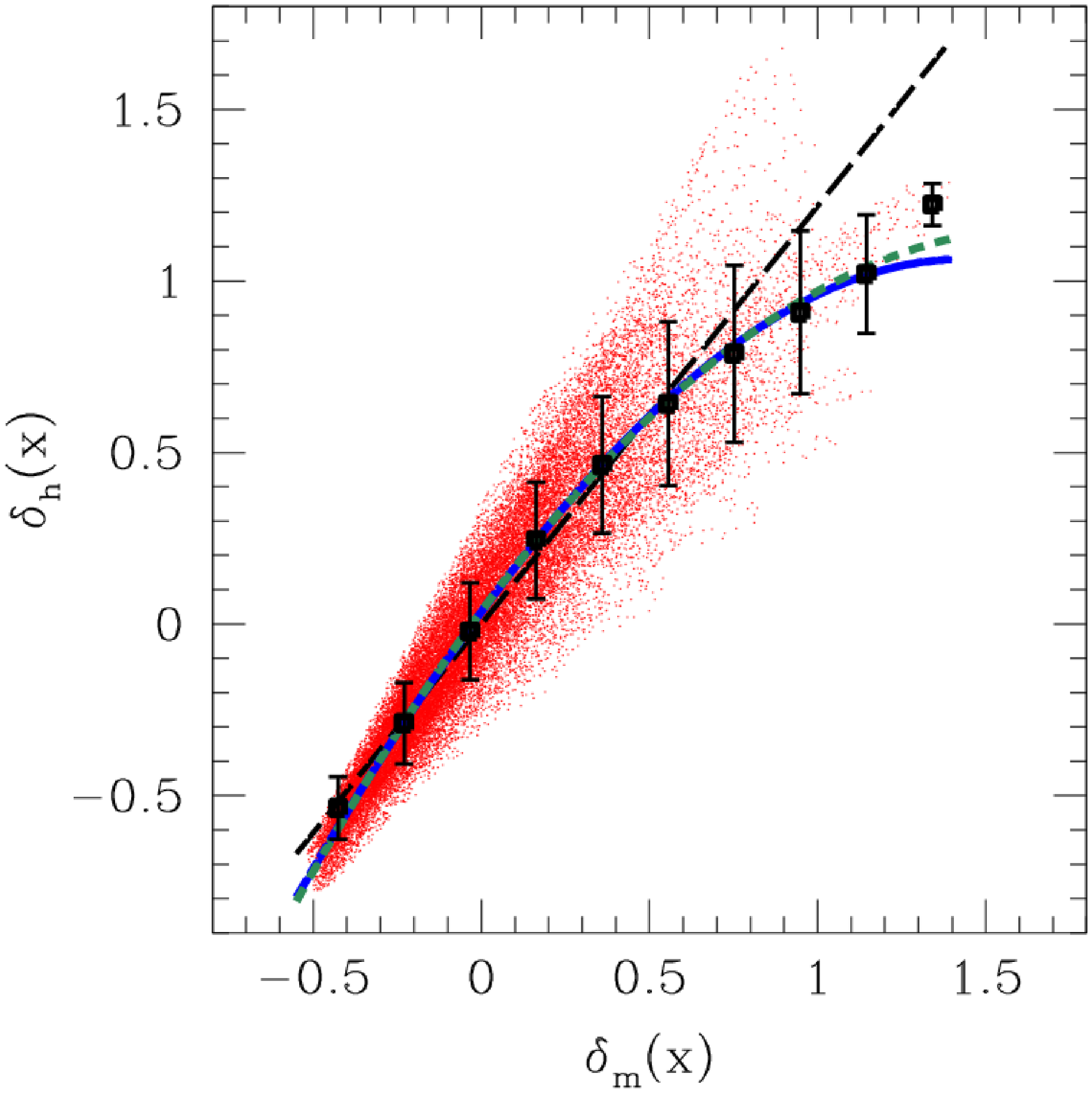}
\includegraphics[scale=0.35]{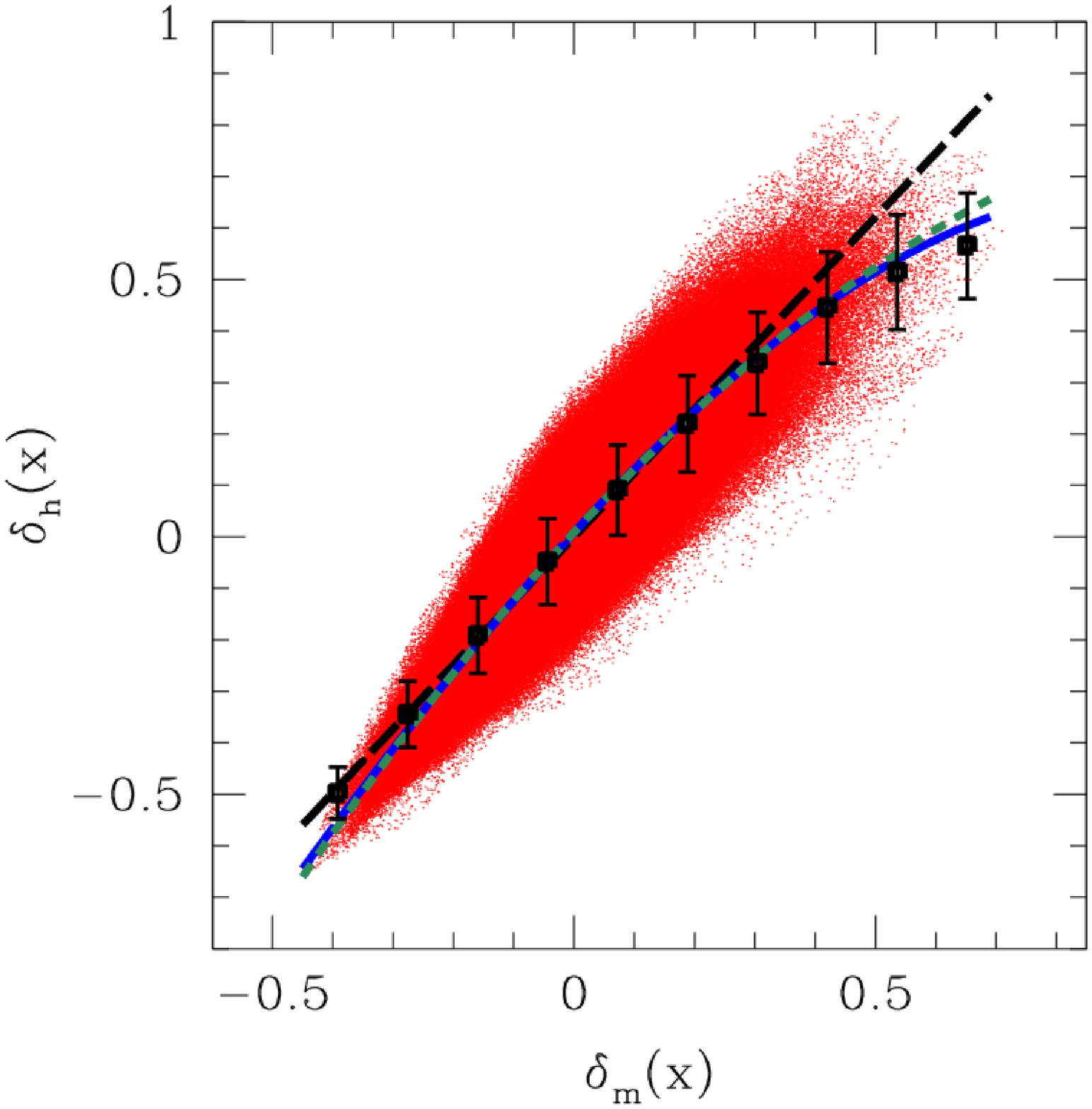}\\
\includegraphics[scale=0.35]{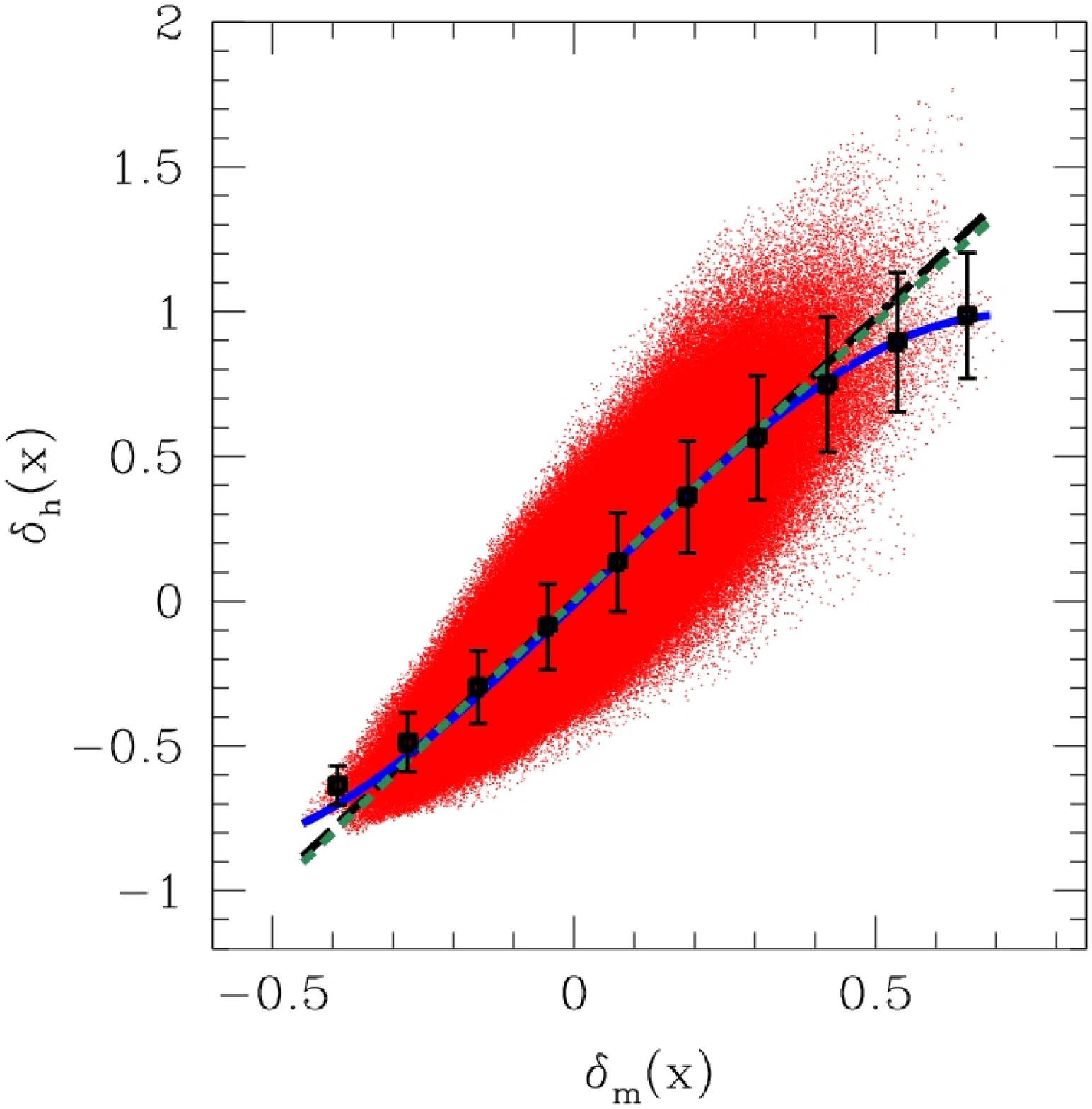}
\includegraphics[scale=0.35]{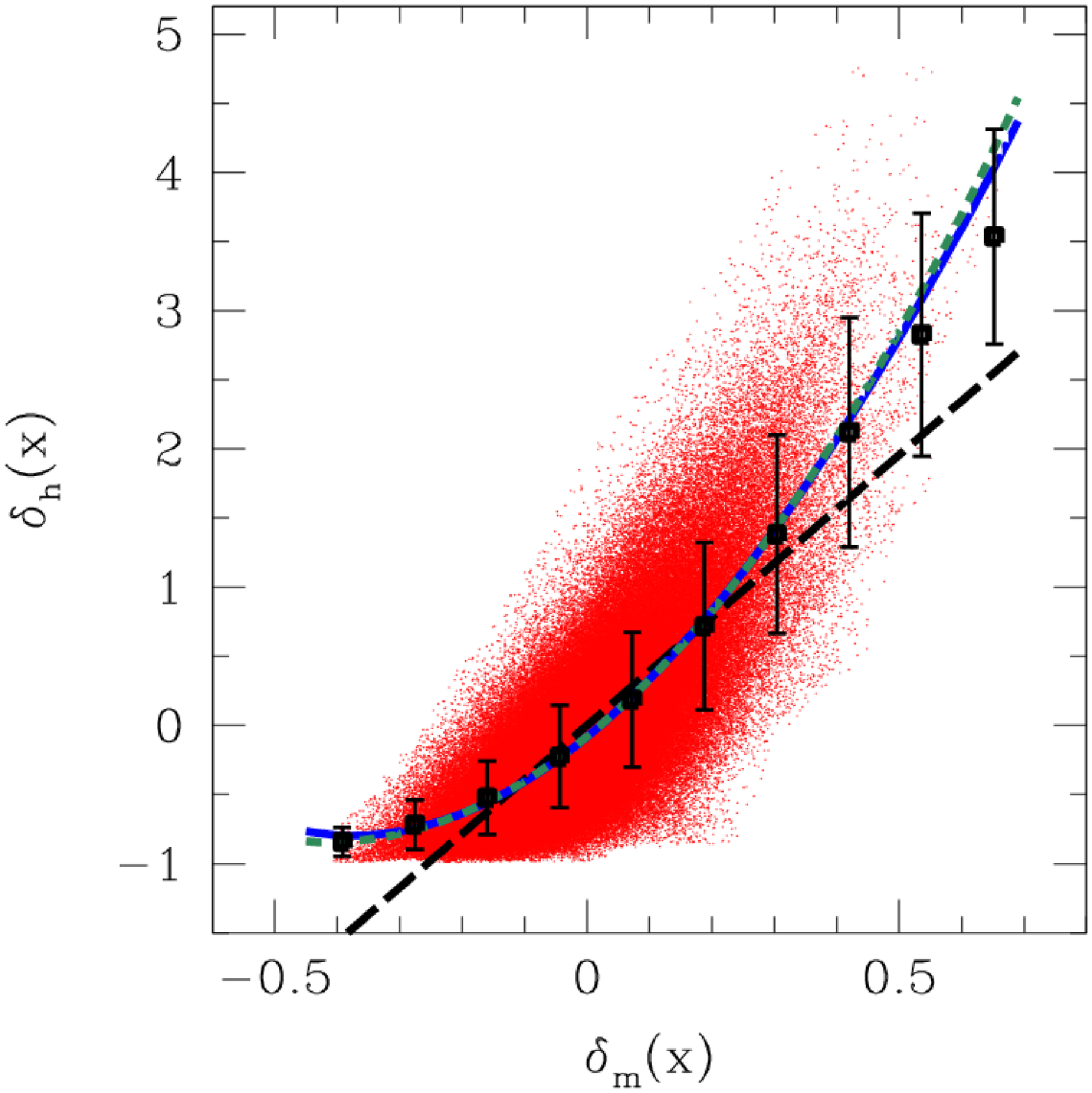}
\caption{Scatter plot of $\delta_{\mathrm{h}}^{\mathrm{s}}(\mathbf{x})$ vs. $\delta_{\mathrm{m}}^{\mathrm{s}}(\mathbf{x})$. Top row: mass bins I and II ($R=12\ \mathrm{Mpc}/h$), middle row: bins III ($R=12\ \mathrm{Mpc}/h$) and IV ($R=28\ \mathrm{Mpc}/h$), bottom row: bins V and VI ($R=28\ \mathrm{Mpc}/h$). The lines show the polynomials of different order fitted to determine the bias parameters $b_i$ (black long-dashed: linear, green short-dashed: second order, solid blue: third order). The black squares with errorbars show the mean of $\delta_{\mathrm{h}}^{\mathrm{s}}$ in bins of $\delta_{\mathrm{m}}^{\mathrm{s}}$ and the 1-$\sigma$ scatter around it as visual guidance. Note that, for readability, we only show one of every eight points used in the fitting procedure.}
\label{fig:fitsb12}
\end{figure*}

Our fitting process allows for three free parameters: $b_1$, $b_2$ and $b_3$ ($b_0$ is not a free parameter and always very close to 0 so it can be neglected). We also fit the point cloud with fewer bias parameters and use the Akaike information criterion\footnote{Basically a $\chi^2$ method which penalizes extra parameters.} (AIC) to determine the best fit without overfitting the data with too many parameters. We define
\begin{align}
\delta_{\mathrm{h},\mathrm{f}_1}(\mathbf{x}) 
\equiv &\  b_1 \delta_{\mathrm{m}}^{\mathrm{s}}(\mathbf{x}), \notag \\
\delta_{\mathrm{h},\mathrm{f}_2}(\mathbf{x}) 
\equiv &\ b_1 \delta_{\mathrm{m}}^{\mathrm{s}}(\mathbf{x})+\frac{b_2}{2} \left(\delta_{\mathrm{m}}^{\mathrm{s}}(\mathbf{x}) \right)^2, \nonumber\\
\delta_{\mathrm{h},\mathrm{f}_3}(\mathbf{x}) 
\equiv &\ b_1 \delta_{\mathrm{m}}^{\mathrm{s}}(\mathbf{x}) +\frac{b_2}{2} \left( \delta_{\mathrm{m}}^{\mathrm{s}}(\mathbf{x}) \right)^2  + \frac{b_3}{6} \left( \delta_{\mathrm{m}}^{\mathrm{s}}(\mathbf{x}) \right)^3,
\label{fdm}
\end{align}
which describes the parameterisation used for fitting and the notation to denote the halo density contrast $\delta_{\mathrm{h},\mathrm{f}_i}(\mathbf{x})$ obtained from the $i$-th order fit. Note that $\delta_{\mathrm{h},\mathrm{f}_i}(\mathbf{x})$ is not written with the superscript s due to the unsmoothing effect discussed in the previous paragraph. The errors on the parameters are calculated using the jackknife method with 8 subsamples of the density contrast fields. Scatter plots of $\delta_{\mathrm{h}}^{\mathrm{s}}(\mathbf{x})$ vs. $\delta_{\mathrm{m}}^{\mathrm{s}}(\mathbf{x})$ and the fitted polynomials are shown in Fig. \ref{fig:fitsb12} for all six mass bins. The mean and $1-\sigma$ scatter for $\delta_{\mathrm{h}}^{\mathrm{s}}$ in bins of $\delta_{\mathrm{m}}^{\mathrm{s}}$  are indicated by points with errorbars. One can see that the local bias model seems to be a rather good approximation, although there is some scatter in the relation. A possible source of this could be shot noise from sampling discrete haloes. Assuming Poissonian shot noise, the amplitude can be estimated by calculating the number of haloes $N_{\mathrm{s}}= \rho_{\mathrm{h}}\, V_{\mathrm{s}} $ that are in each smoothing volume $V_{\mathrm{s}}=6 \pi^2 R^3$, and the mean number of haloes in that mass bin, $\overline{N}$, from the third column in Table \ref{tab:bins}:
\begin{equation}
\Delta \delta_{\mathrm{h}}^{\mathrm{s}}= \frac{\Delta N_{\mathrm{s}}}{ \overline{N}} = \frac{\sqrt{ \overline{N} \left( \delta_{\mathrm{h}}^{\mathrm{s}} +1\right)}} { \overline{N}}.
\end{equation}
where $\Delta N_{\mathrm{s}}= \sqrt{N_{\mathrm{s}}}$ (Poissonian noise) was used in the second step. For mass bin I with $R=12\ \mathrm{Mpc}/h$, $|\delta_{\mathrm{h}}^{\mathrm{s}}| \leq 0.8$. The maximum error from shot noise would be $\Delta (\delta_{\mathrm{h}}^{\mathrm{s}}=0.8) \approx 0.03$. But one can see from Fig. \ref{fig:fitsb12} that the scatter does not increase as $|\delta_{\mathrm{h}}^{\mathrm{s}}|$ increases, and that it is much larger than the maximum shot noise error. This means that there is an \textit{intrinsic scatter} around the local bias model, i.e. the bias is not deterministic (see also \citealt{2001MNRAS.320..289S}). For the higher mass bins with fewer haloes (right panel), the effect of shot noise is more prominent. There have been suggestions in the literature how to deal with stochasticity (e.g. \citealt*{Dekel:1998eq}; \citealt*{2011MNRAS.412..995C}), but we will not include this in our modelling. The AIC values can be converted into a relative probability $w_i$ for each model (see e.g. the appendix of \citealt*{2006MNRAS.371.1824P}), with $\sum_i w_i=1$. Independent of halo masses and smoothing scale $R$, the third-order bias model always provides the best fit, with  $w_3 \approx 1$ while $w_1,w_2$ are lower by several orders of magnitudes. This means that although the second- and third-order fits in Fig. \ref{fig:fitsb12} differ only in regions with relatively few points, the AIC criterion clearly favours a model with 3 bias parameters.
\begin{figure*}
\centering
\includegraphics[scale=0.3]{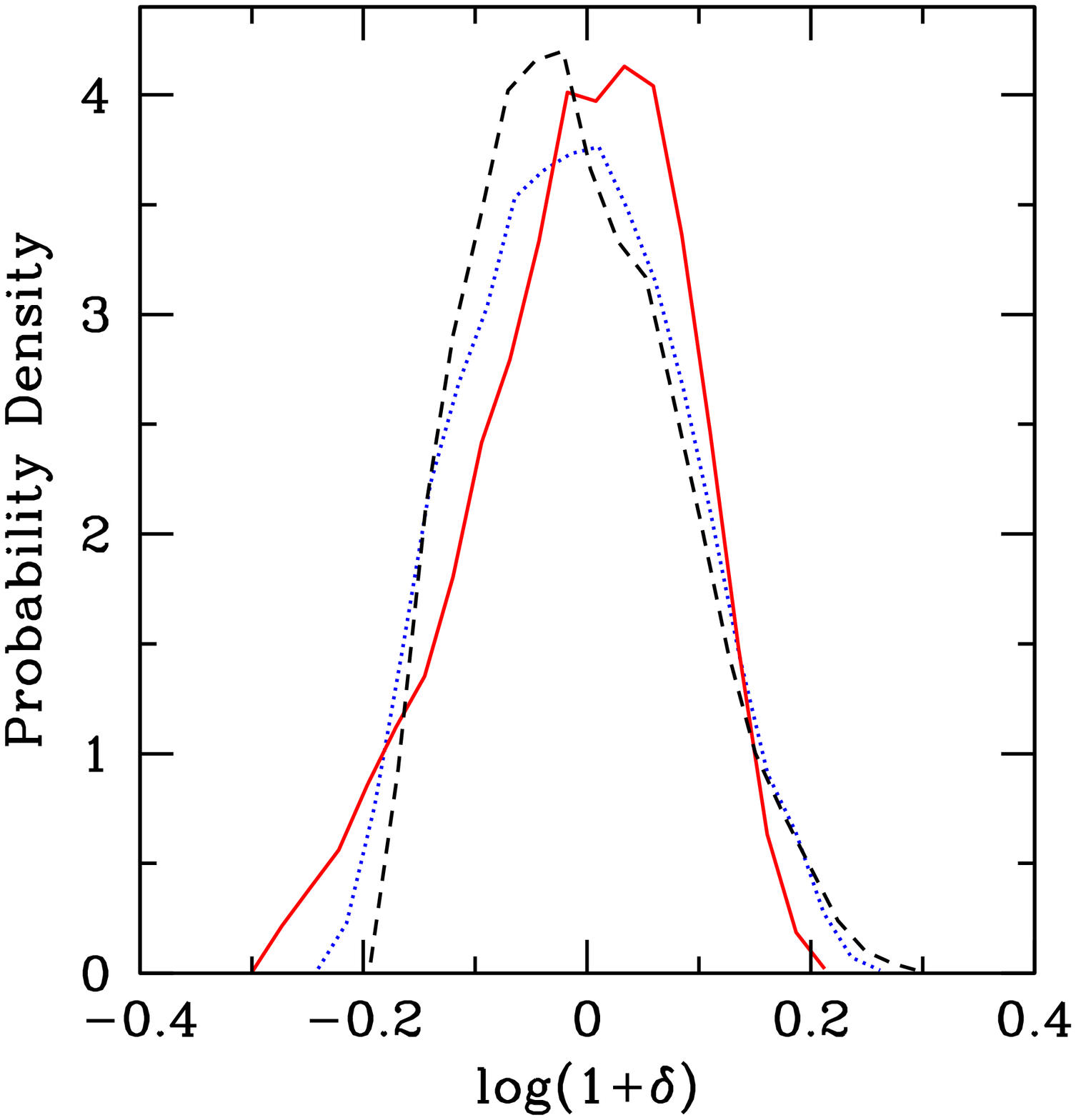}
\includegraphics[scale=0.3]{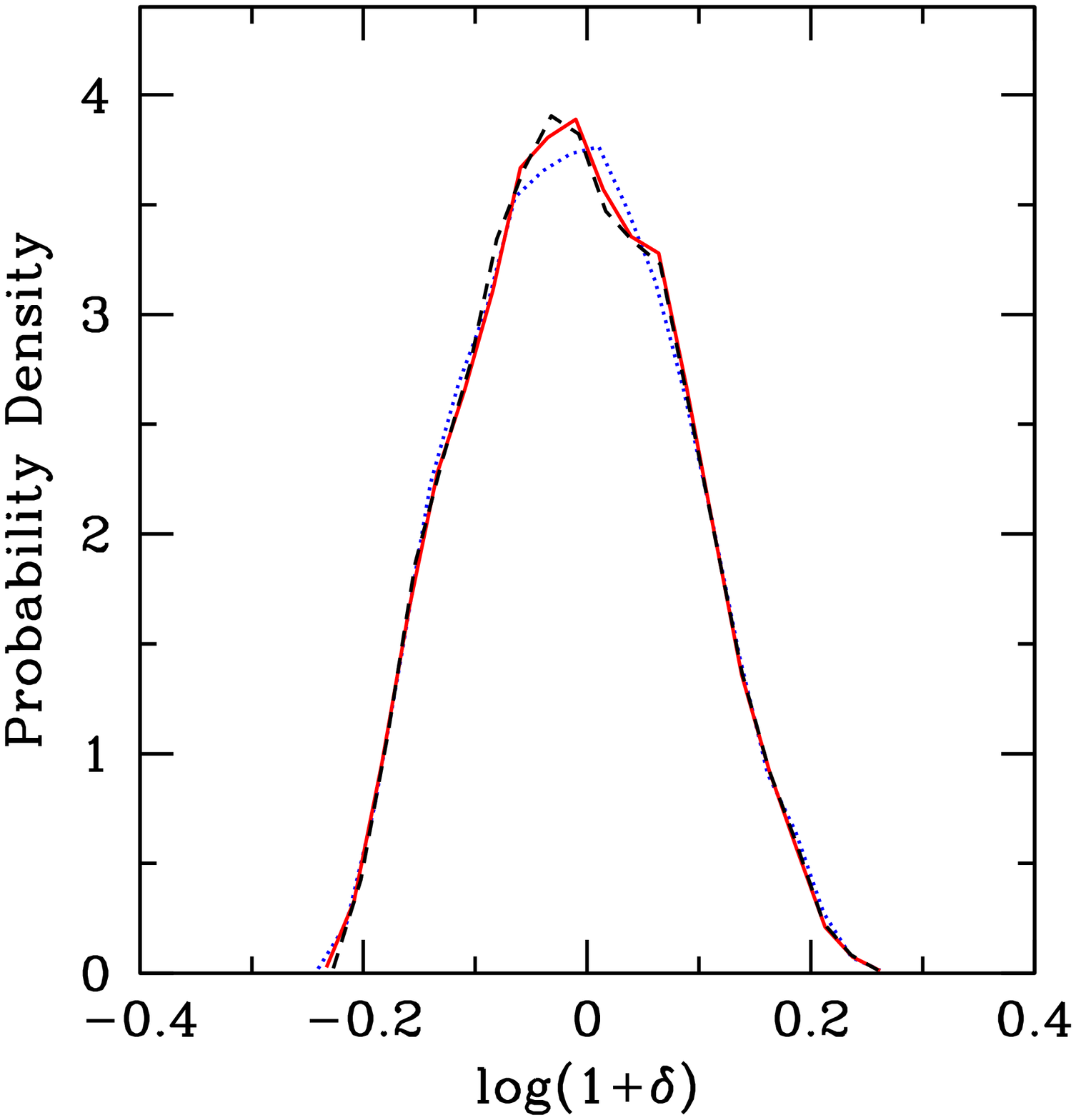}
\caption{Left: Probability distribution function for $\delta_{\mathrm{h}}(\mathbf{x})$ (dotted blue) compared to $\delta_{\mathrm{h},f_1}(\mathbf{x})$ (black dashed) and $\delta^{\mathrm{SPT}}_{\mathrm{h},f_1}(\mathbf{x})$ (red solid). Right: Probability distribution function for $\delta_{\mathrm{h}}(\mathbf{x})$ (dotted blue) compared to $\delta_{\mathrm{h},f_3}(\mathbf{x})$ (black dashed) and $\delta^{\mathrm{SPT}}_{\mathrm{h},f_3}(\mathbf{x})$ (red solid). Both panels show mass bin I with $R=12\ \mathrm{Mpc}/h$.}
\label{fig:pdf_haloes}
\end{figure*}

In the previous sections, we showed that $\delta_{\mathrm{SPT}}^{\mathrm{s}} \approx \delta_{\mathrm{m}}^{\mathrm{s}}$. So instead of fitting to the simulation matter density contrast, we can also obtain the bias parameters by fitting $\delta_{\mathrm{h}}^{\mathrm{s}}(\mathbf{x})$ vs. $\delta_{\mathrm{SPT}}^{\mathrm{s}}(\mathbf{x})$. Similar to Eq. (\ref{fdm}), we define
\begin{align}
\delta_{\mathrm{h},\mathrm{f}_1}^{\mathrm{SPT}}(\mathbf{x})
\equiv &\   \bar{b}_1 \delta_{\mathrm{1}}^{\mathrm{s}}(\mathbf{x}) \nonumber, \\
\delta_{\mathrm{h},\mathrm{f}_2}^{\mathrm{SPT}}(\mathbf{x}) 
\equiv &\ \bar{b}_1 \left[ \delta_{\mathrm{1}}^{\mathrm{s}}(\mathbf{x})+\delta_{\mathrm{2}}^{\mathrm{s}}(\mathbf{x})\right]+\frac{\bar{b}_2}{2} \left(\delta_{\mathrm{1}}^{\mathrm{s}}(\mathbf{x}) \right)^2, \nonumber \\
\delta_{\mathrm{h},\mathrm{f}_3}^{\mathrm{SPT}}(\mathbf{x}) 
\equiv &\ \bar{b}_1 \left[ \delta_{\mathrm{1}}^{\mathrm{s}}(\mathbf{x})+\delta_{\mathrm{2}}^{\mathrm{s}}(\mathbf{x})+\delta_{\mathrm{3}}^{\mathrm{s}}(\mathbf{x}) \right] +\frac{\bar{b}_2}{2} \left[ \left(\delta_{\mathrm{1}}^{\mathrm{s}}(\mathbf{x}) \right)^2 \right. \nonumber \\
&  +2\delta_{\mathrm{1}}^{\mathrm{s}}(\mathbf{x})\delta_{\mathrm{2}}^{\mathrm{s}}(\mathbf{x}) \big]  + \frac{\bar{b}_3}{6} \left( \delta_{\mathrm{1}}^{\mathrm{s}}(\mathbf{x}) \right)^3.
\label{fpt}
\end{align}
The bar is used to point out that $b_i$ and $\bar{b}_i$ are generally different. This is not particularly surprising because while we showed before that $\delta_{\mathrm{m}}^{\mathrm{s}} \approx \delta_{\mathrm{SPT}}^{\mathrm{s}}$, the density contrast we fit to for the $\bar{b}_i$ is truncated at the order of the fit in $\delta_1$. Considering the third-order fit, one therefore expects the largest deviation between the fit parameters to occur between $b_3$ and $\bar{b}_3$, which is indeed the case except for the high mass bin V (see Table \ref{tab:biasdmb1}). This truncation is consistent with previous works, and using the full SPT density contrast in the second- and third-order fits does not improve the agreement between $b_i$ and $\bar{b}_i$. The errors on $\bar{b}_i$ are also jackknife errors from 8 subsamples, and again the third-order fit is always preferred by the AIC. Note that the bias parameters also depend on the smoothing scale $R$. Both effects can be seen in Table \ref{tab:biasdmb1} for mass bins I, III and V. We do not show the values of the fitted bias parameters for all possible combinations of mass bins and smoothing scales, but we comment on their mass dependence for a specific smoothing scale in section \ref{ssec:massdep}. The fact that the allowed range for $b_3$ does not include zero in most cases is also an indication that the third-order bias parameter is required for the fit.
\begin{table}
\centering
\caption{Bias parameters for both third-order fits for different mass bins and smoothing scales $R\ [\mathrm{Mpc}/h]$.}
\begin{tabular}{|c|c|c|c|c|}
\hline & Mass bin & $R$ & $\delta_{\mathrm{h}}^{\mathrm{s}}$ vs. $\delta_{\mathrm{m}}^{\mathrm{s}}$ & $\delta_{\mathrm{h}}^{\mathrm{s}}$ vs. $\delta_{\mathrm{SPT}}^{\mathrm{s}}$  \\ 
\hline 
$b_1$& I & 8  & $\ \ 0.747 \pm 0.006 $ & $\ \ 0.666 \pm 0.008 $ \\ 
$b_2$& I & 8  & $ -0.341 \pm 0.022 $ & $ -0.554 \pm 0.021 $ \\ 
$b_3$& I & 8  & $\ \ 0.171 \pm 0.034$ & $\ \ 1.542 \pm 0.048 $ \\
$b_1$& I & 12 & $\ \  0.719 \pm 0.009$ & $\ \ 0.711 \pm 0.010$ \\ 
$b_2$& I & 12 & $-0.300 \pm 0.034 $ & $-0.438 \pm 0.017$  \\ 
$b_3$& I & 12 & $\ \ 0.171 \pm 0.125 $ & $\ \ 0.620 \pm 0.111$  \\ 
$b_1$& III & 12 &  $\ \ 1.333 \pm 0.035 $ & $\ \ 1.323 \pm 0.034 $ \\ 
$b_2$& III & 12 & $ -0.716 \pm 0.140 $ & $ -1.086 \pm 0.093 $ \\ 
$b_3$& III & 12 & $ -0.288 \pm 0.612 $ & $\ \ 0.066 \pm 0.532 $ \\ 
$b_1$& V & 28  & $\ \ 2.039 \pm 0.023 $ & $\ \ 2.037 \pm 0.023 $ \\ 
$b_2$& V & 28  & $\ \ 0.270 \pm 0.124 $ & $ -0.346 \pm 0.136 $ \\ 
$b_3$& V & 28  & $ -8.682 \pm 1.476 $ & $ -8.556 \pm 1.446 $ \\
\hline
\end{tabular}
\label{tab:biasdmb1}
\end{table}
\subsubsection{Testing the Local Bias Assumption}
\label{sssec:test}
\begin{table}
\centering
\caption{Correlation coefficients for the halo density contrast for a selection of mass bins and smoothing scales $R\ [\mathrm{Mpc}/h]$.}
\begin{tabular}{|c|c|c|c|c|}
\hline Mass bin & $R$ & Pair & $c$ & $c_{\mathrm{Sp}}$ \\ 
\hline 
I & 12 & $\langle \delta_{\mathrm{h}}^{\mathrm{s}}\ \delta_{\mathrm{h},\mathrm{f}_1} \rangle$ & 0.987 & 0.989 \\ 
I & 12 & $\langle \delta_{\mathrm{h}}^{\mathrm{s}}\ \delta_{\mathrm{h},\mathrm{f}_2} \rangle$ & 0.990 & 0.989 \\ 
I & 12 & $\langle \delta_{\mathrm{h}}^{\mathrm{s}}\ \delta_{\mathrm{h},\mathrm{f}_3} \rangle$ & 0.990 & 0.989 \\ 
I & 12 & $\langle \delta_{\mathrm{h}}^{\mathrm{s}}\ \delta^{\mathrm{spt}}_{\mathrm{h},\mathrm{f}_1} \rangle$ & 0.951 & 0.961 \\ 
I & 12 & $\langle \delta_{\mathrm{h}}^{\mathrm{s}}\ \delta^{\mathrm{spt}}_{\mathrm{h},\mathrm{f}_2} \rangle$ & 0.985 & 0.986 \\ 
I & 12 & $\langle \delta_{\mathrm{h}}^{\mathrm{s}}\ \delta^{\mathrm{spt}}_{\mathrm{h},\mathrm{f}_3} \rangle$ & 0.989 & 0.989 \\
III & 12 & $\langle \delta_{\mathrm{h}}^{\mathrm{s}}\ \delta_{\mathrm{h},\mathrm{f}_1} \rangle$ & 0.934 & 0.956  \\ 
III & 12 & $\langle \delta_{\mathrm{h}}^{\mathrm{s}}\ \delta_{\mathrm{h},\mathrm{f}_2} \rangle$ & 0.943 & 0.956 \\ 
III & 12 & $\langle \delta_{\mathrm{h}}^{\mathrm{s}}\ \delta_{\mathrm{h},\mathrm{f}_3} \rangle$ & 0.043 & 0.956 \\ 
III & 12 & $\langle \delta_{\mathrm{h}}^{\mathrm{s}}\ \delta^{\mathrm{spt}}_{\mathrm{h},\mathrm{f}_1} \rangle$ & 0.902 & 0.922 \\ 
III & 12 & $\langle \delta_{\mathrm{h}}^{\mathrm{s}}\ \delta^{\mathrm{spt}}_{\mathrm{h},\mathrm{f}_2} \rangle$ & 0.937 & 0.953 \\ 
III & 12 & $\langle \delta_{\mathrm{h}}^{\mathrm{s}}\ \delta^{\mathrm{spt}}_{\mathrm{h},\mathrm{f}_3} \rangle$ & 0.944 & 0.957 \\
V & 28 & $\langle \delta_{\mathrm{h}}^{\mathrm{s}}\ \delta_{\mathrm{h},\mathrm{f}_1} \rangle$ & 0.876 & 0.875 \\ 
V & 28 & $\langle \delta_{\mathrm{h}}^{\mathrm{s}}\ \delta_{\mathrm{h},\mathrm{f}_2} \rangle$ & 0.876 & 0.875 \\ 
V & 28 & $\langle \delta_{\mathrm{h}}^{\mathrm{s}}\ \delta_{\mathrm{h},\mathrm{f}_3} \rangle$ & 0.877 & 0.875 \\ 
V & 28 & $\langle \delta_{\mathrm{h}}^{\mathrm{s}}\ \delta^{\mathrm{spt}}_{\mathrm{h},\mathrm{f}_1} \rangle$ & 0.866 & 0.869 \\ 
V & 28 & $\langle \delta_{\mathrm{h}}^{\mathrm{s}}\ \delta^{\mathrm{spt}}_{\mathrm{h},\mathrm{f}_2} \rangle$ & 0.876 & 0.875 \\ 
V & 28 & $\langle \delta_{\mathrm{h}}^{\mathrm{s}}\ \delta^{\mathrm{spt}}_{\mathrm{h},\mathrm{f}_3} \rangle$ & 0.877 & 0.875 \\
\hline 
\end{tabular}
\label{tab:corrsptb1}
\end{table}

\begin{figure*}
\centering
\includegraphics[scale=0.3]{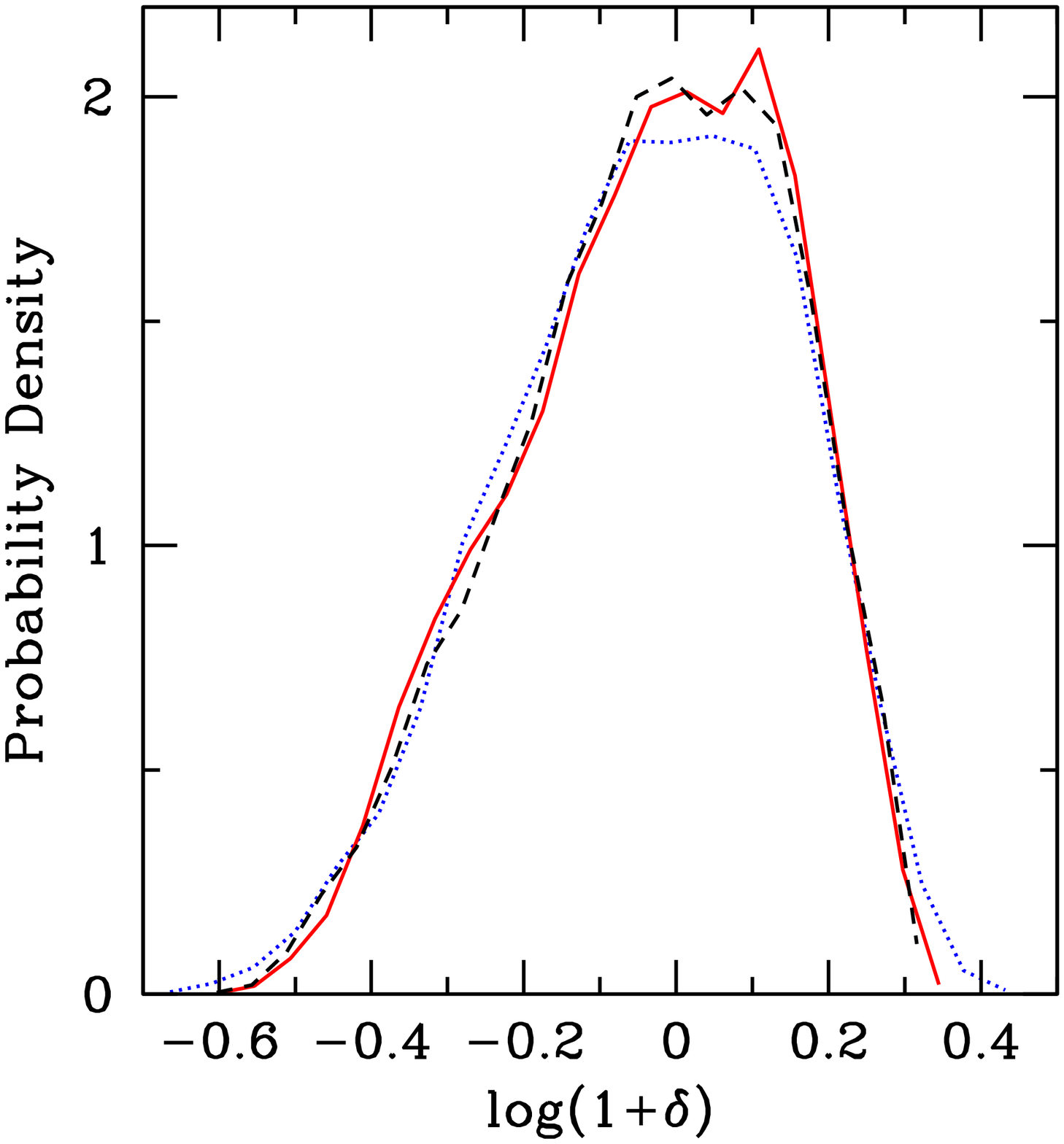}
\includegraphics[scale=0.3]{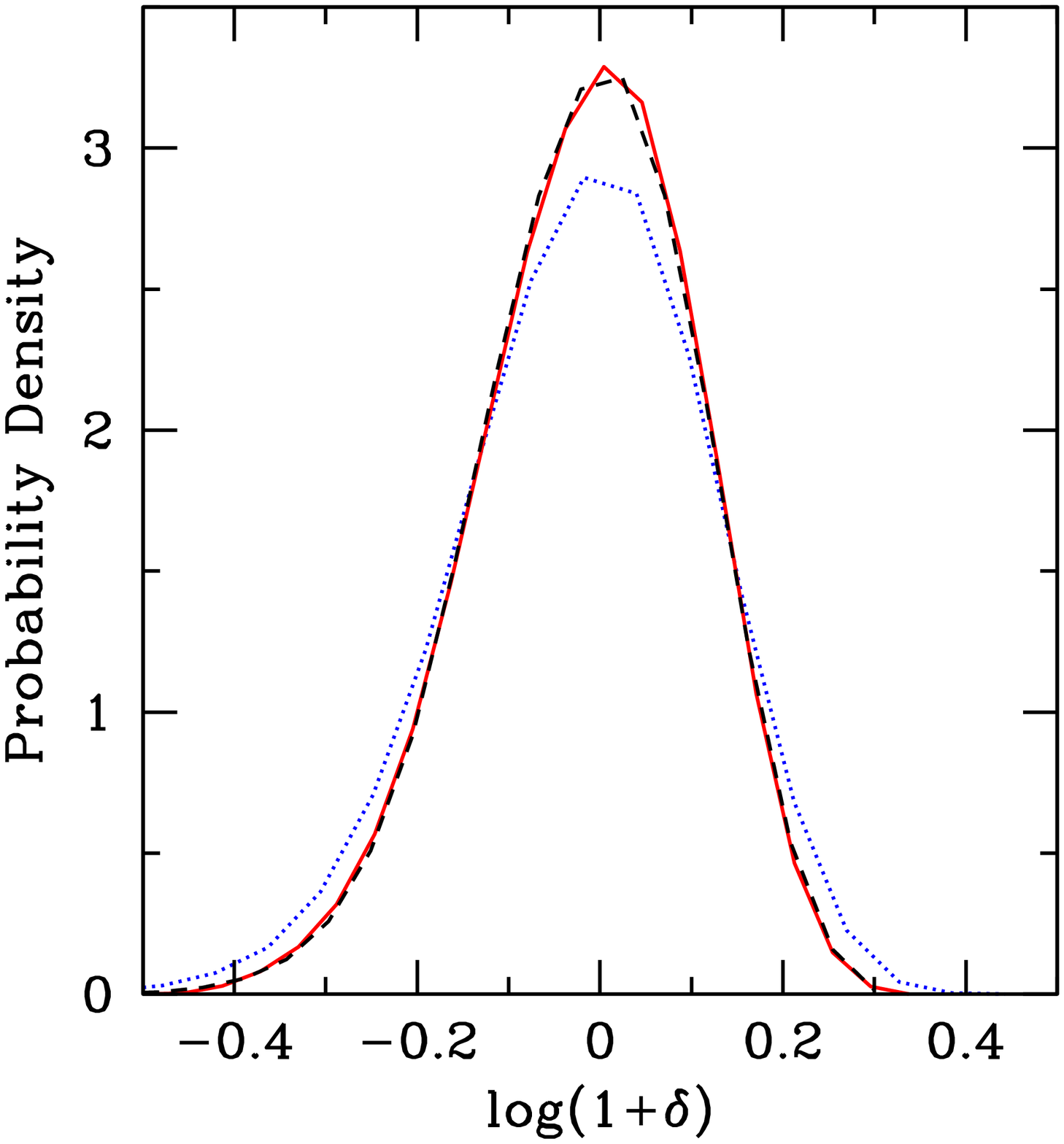}
\caption{Probability distribution function for $\delta_{\mathrm{h}}(\mathbf{x})$ (dotted blue) compared to $\delta_{\mathrm{h},f_3}(\mathbf{x})$ (black dashed) and $\delta^{\mathrm{SPT}}_{\mathrm{h},f_3}(\mathbf{x})$ (red solid). Left: Medium mass bin III with $R=12\ \mathrm{Mpc}/h$. Right: High mass bin V with $R=28\ \mathrm{Mpc}/h$.}
\label{fig:pdf_haloesb4}
\end{figure*}
\begin{figure*}
\centering
\includegraphics[scale=0.3]{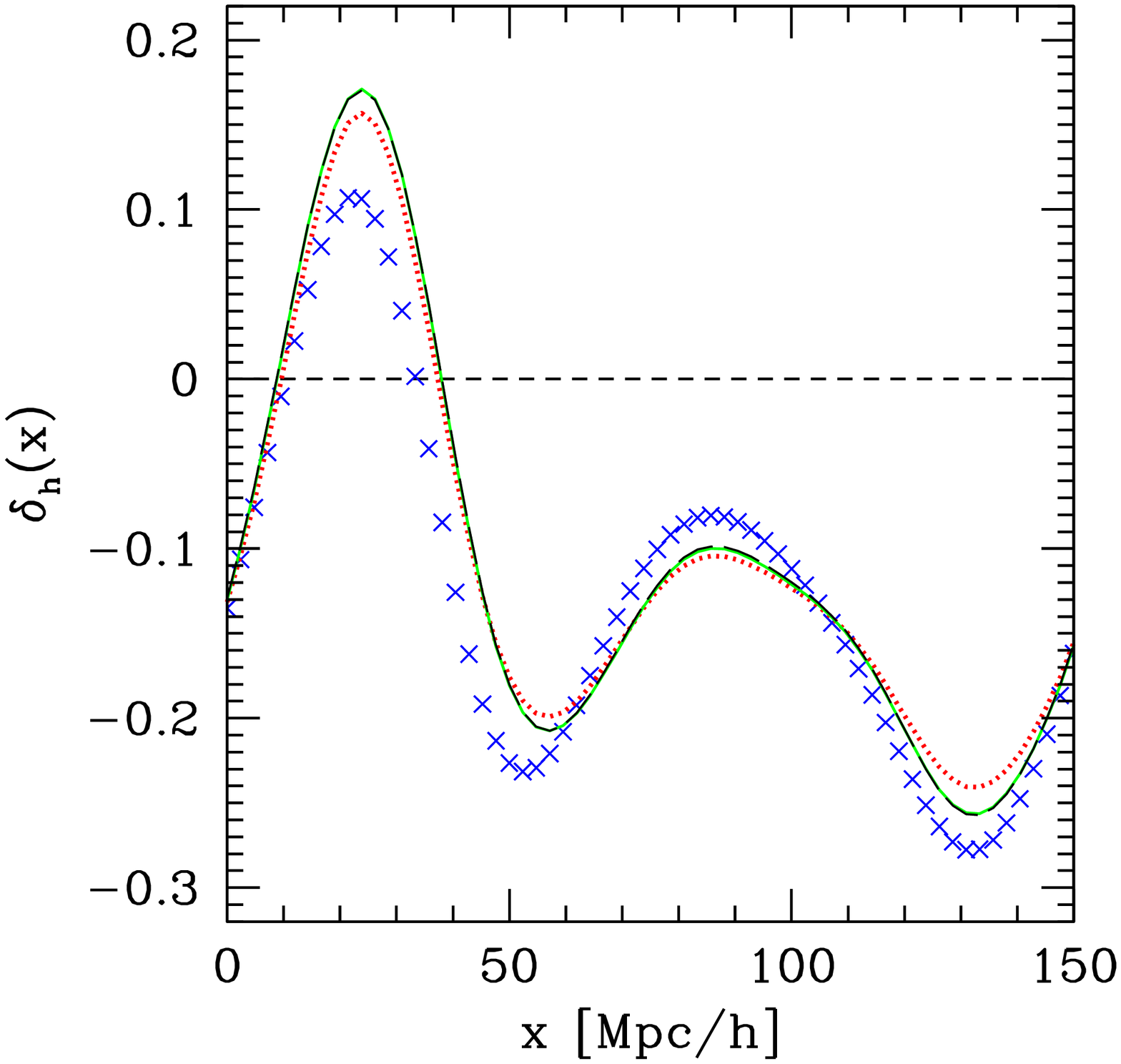}
\includegraphics[scale=0.3]{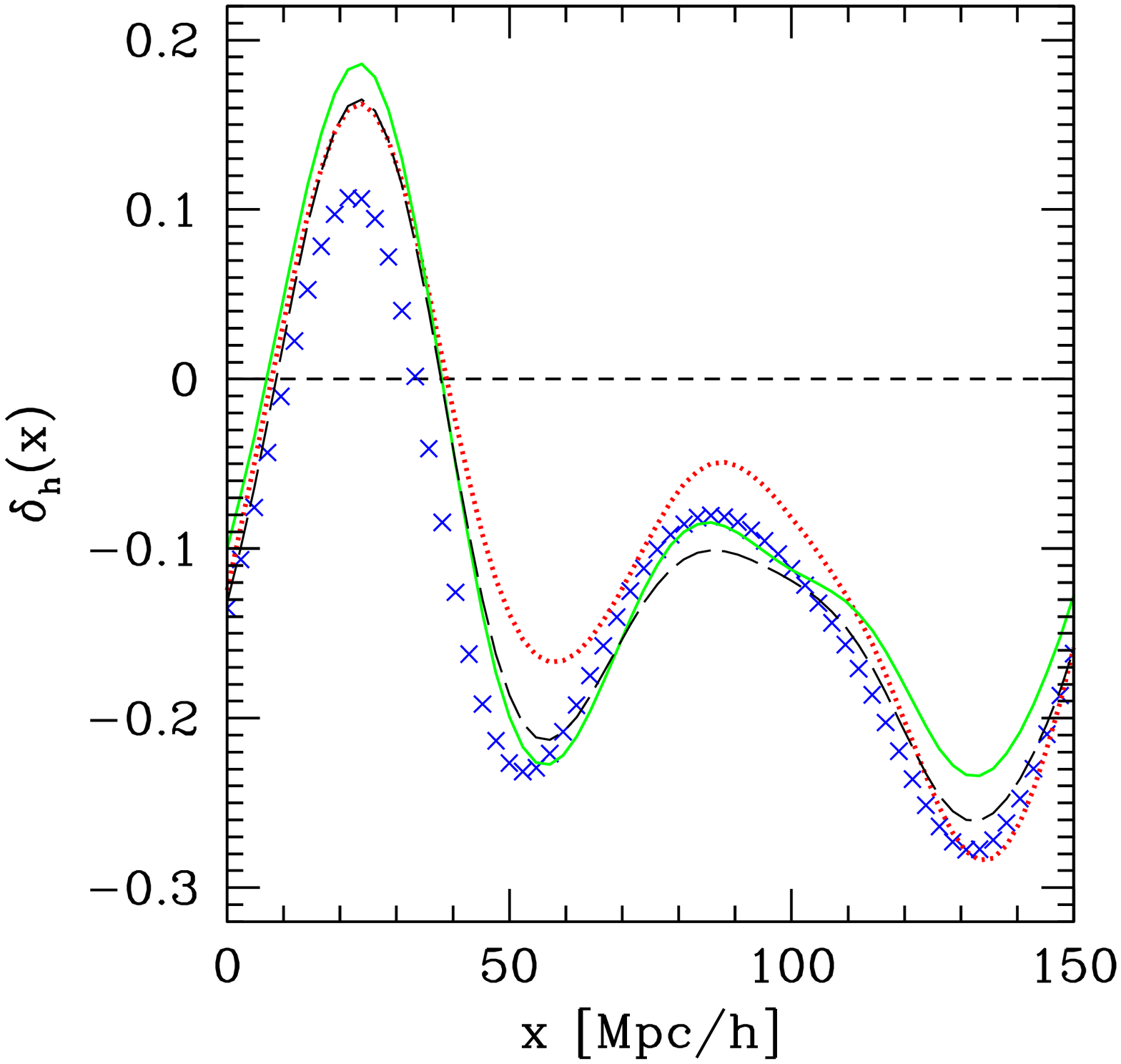}
\caption{Left: Amplitude of the halo density contrast from the simulation (blue crosses) compared to $\delta_{\mathrm{h},\mathrm{f}_i}(\mathbf{x})$. Red dotted: linear fit, green solid: second-order fit, black long-dashed: third-order fit. Right: Same for $\delta_{\mathrm{h},\mathrm{f}_i}^{\mathrm{SPT}}(\mathbf{x})$. Both panels show mass bin I with $R=12\ \mathrm{Mpc}/h$. Note that this is the same region of the density field as in the left panel of Fig. \ref{fig:dxptvsdxm_high}.}
\label{fig:dhvsfdmdm_lowdens}
\end{figure*}
\begin{figure*}
\centering
\includegraphics[scale=0.3]{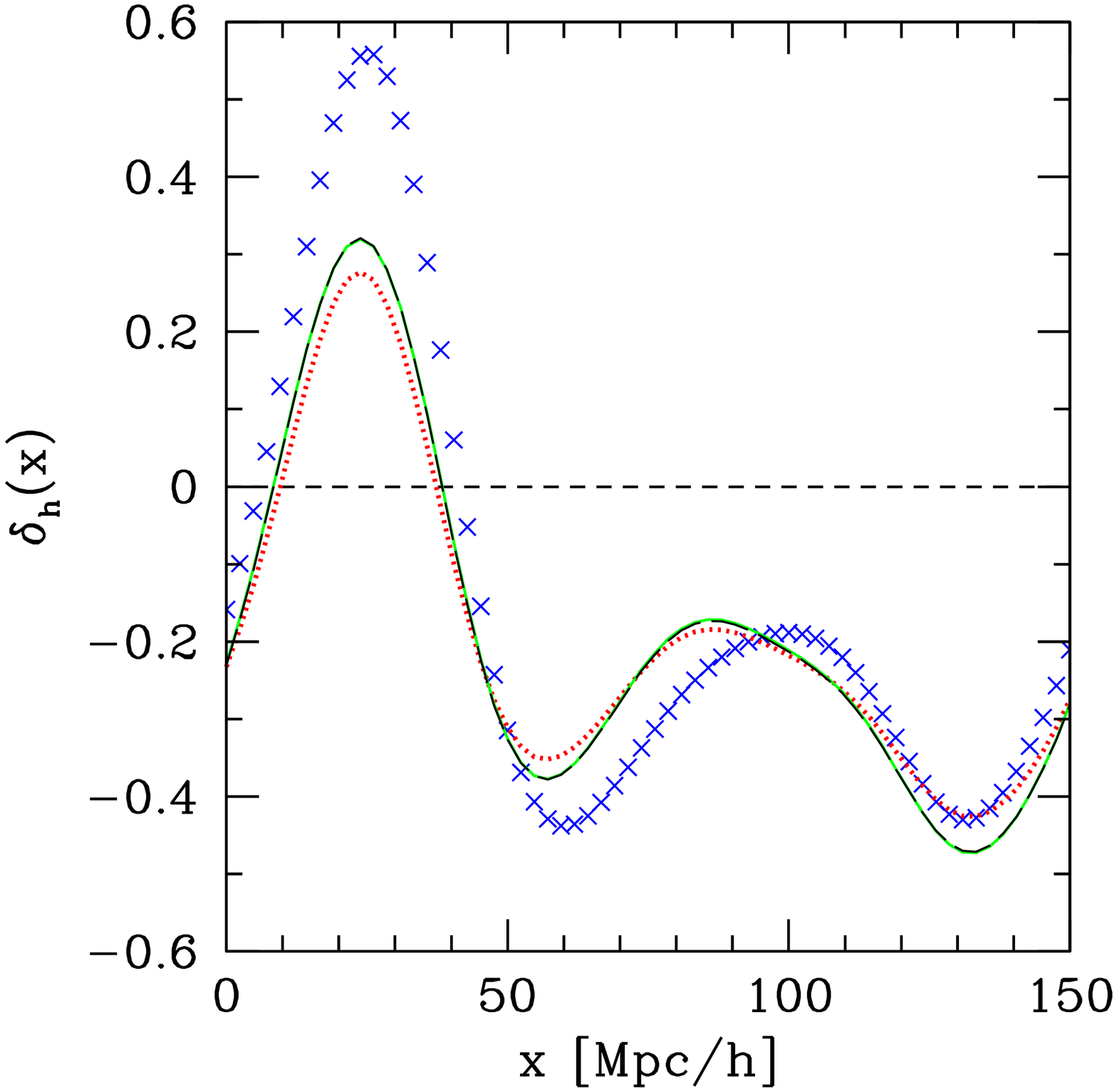}
\includegraphics[scale=0.3]{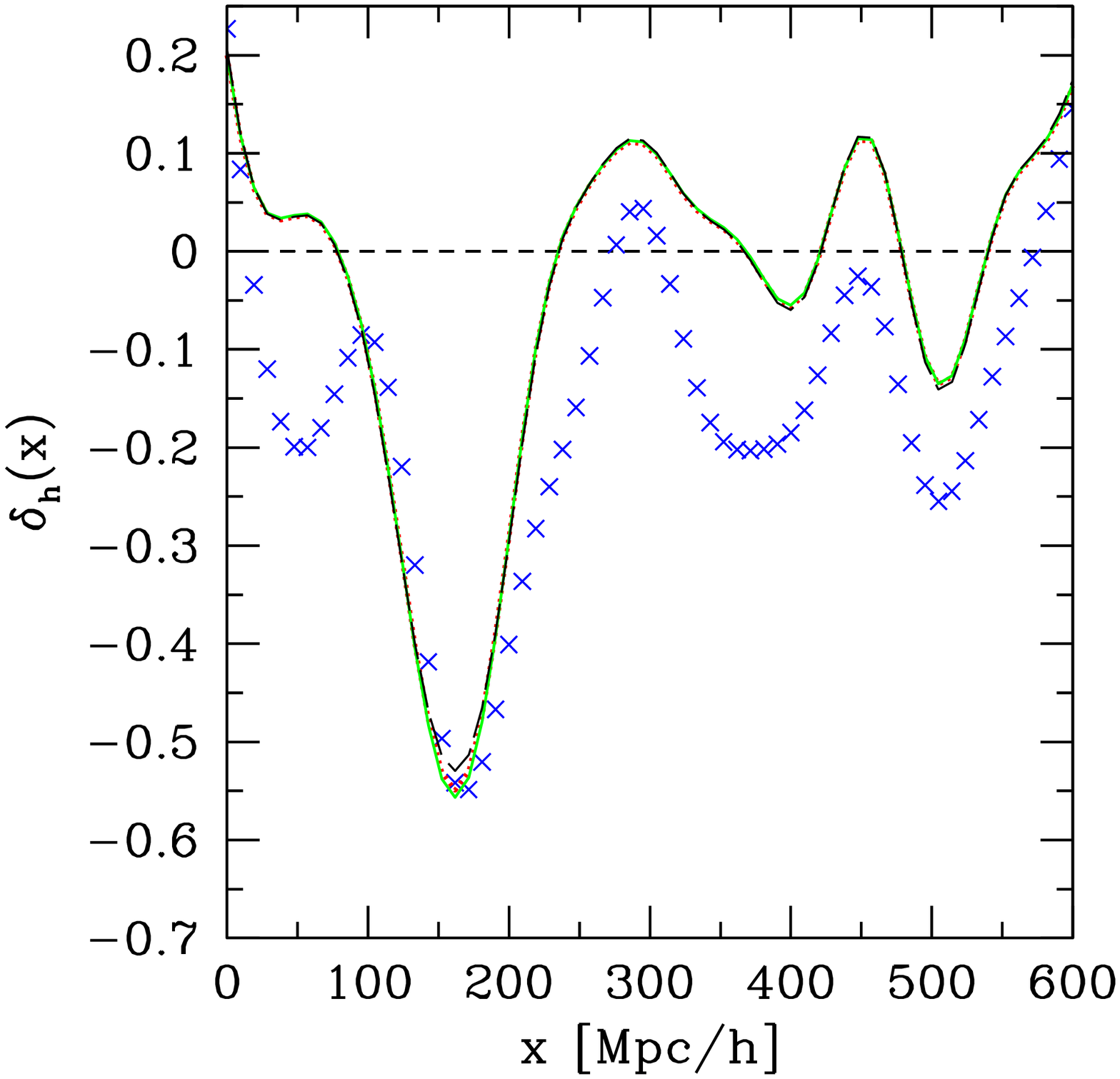}
\caption{Amplitude of the halo density contrast from the simulation (blue crosses) compared to $\delta_{\mathrm{h},\mathrm{f}_i}(\mathbf{x})$. Red dotted: linear fit, green solid: second-order fit, black long-dashed: third-order fit. Left: Medium mass bin III with $R=12\ \mathrm{Mpc}/h$. (Same region as in Fig. \ref{fig:dhvsfdmdm_lowdens}). Right: High mass bin V with $R=28\ \mathrm{Mpc}/h$.}
\label{fig:dhvsfdmdm_otherbin}
\end{figure*}

Now we can compare the fitted halo density contrast $\delta_{\mathrm{h},\mathrm{f}_i}(\mathbf{x})$  to the true halo density contrast from the simulations $\delta_{\mathrm{h}}^{\mathrm{s}}(\mathbf{x})$. Fig. \ref{fig:pdf_haloes} shows the probability distribution function of the fitted halo density contrast compared to the simulation. The left panel shows the results using a linear fit, i.e. only one bias parameter $b_1$, fitting to $\delta_{\mathrm{m}}^{\mathrm{s}}$ and $\delta_1^{\mathrm{s}}$ respectively. Neither fit gives a satisfactory halo density contrast, and the PDFs look quite different. The right panel shows the same but this time using the third-order fit with all three bias parameters. Here, both fits give PDFs which are almost indistinguishable, but the agreement with the simulated halo density contrast is still not perfect. This discrepancy can be seen for all mass bins: Fig. \ref{fig:pdf_haloesb4} shows the PDFs of the third-order fits for mass bins III and V compared to the simulated density contrast. This indicates that the local bias model does not capture all the properties of the halo density field. 

Fig. \ref{fig:dhvsfdmdm_lowdens} shows again two slices of the simulation along one axis, extending over the box length $L=150\ \mathrm{Mpc}/h$. Overplotted are the fits of different order to $\delta_{\mathrm{m}}^{\mathrm{s}}$ (left panel) and to $\delta_{\mathrm{SPT}}^{\mathrm{s}}$ (right panel). In the former case, the different fits behave very similarly. In the latter case, the third-order fit (black long-dashed) seems to be closest to the simulation. It is clear that in both cases, the local bias model does not resemble the simulated halo density field on small scales. We also show the results for a medium and a high mass bin (III and V) in Fig. \ref{fig:dhvsfdmdm_otherbin}, but here we show only the fits to $\delta_{\mathrm{m}}^{\mathrm{s}}$. Also for higher masses and larger smoothing scales, the ELB model is not satisfactory on a point-by-point level, although the PDFs shown before were similar.

We also calculate the linear correlation coefficient $c$ and the Spearman ranked correlation coefficient $c_{\mathrm{Sp}}$ which allows for a general, non-linear correlation between the simulated and fitted $\delta_{\mathrm{h}}(\mathbf{x})$ (Table \ref{tab:corrsptb1}). The values are very close to 1, which shows as well that overall, the local bias assumption is not such a bad model for out data. From now on we will use the third-order fit because it is the model preferred by the AIC, the correlations also support this choice and it is consistent with the order of the SPT calculation.
 \subsubsection{Mass Dependence}
 \label{ssec:massdep}
\begin{figure}
 \centering
 \includegraphics[scale=0.4]{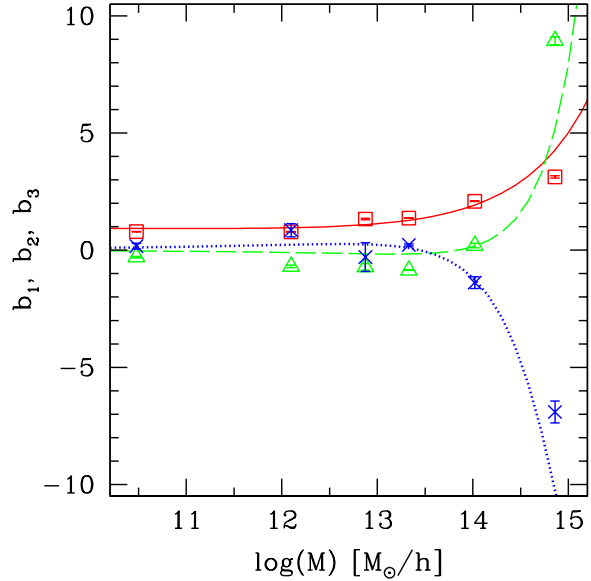}
 \caption{Mass dependence of the fitted bias parameters for $R=12\ \mathrm{Mpc}/h$. Symbols: fit parameters, lines: predictions for the simulated halo mass function. $b_1$ (red squares and solid line), $b_2$ (green triangles and dashed line) and $b_3$ (blue crosses and dotted line).}
 \label{fig:massdep}
\end{figure}

 Fig. \ref{fig:massdep} shows the mass dependence of the bias parameters (fitting $\delta_{\mathrm{h}}^{\mathrm{s}}$ vs. $\delta_{\mathrm{m}}^{\mathrm{s}}$) for the mass bins defined in Table \ref{tab:bins} and $R=12\ \mathrm{Mpc}/h$. Different symbols distinguish the different parameters: red squares for $b_1$, green triangles for $b_2$ and blue crosses for $b_3$. The lines show a theoretical prediction for the bias parameters obtained as follows: The first step is to apply the peak-background-split model to the halo mass function in the simulation (from the fitting formula presented in \citealt{Pillepich08}). The second step uses the spherical collapse model to relate the Lagrangian and Eulerian bias parameters \citep{2010PhRvD..81f3530G}. The two sets of bias parameters show the same trend with halo mass, but they are not in a perfect agreement given the jackknife errors bars (see also \citealt{2010MNRAS.402..589M}). The error in $b_3$ is strongly influenced by shot noise, because there are fewer points in the very high and low density regions which determine the shape of the polynomial.
\subsection{Halo Power Spectra}
\label{sec:halops}
\begin{figure*}
\centering
\includegraphics[scale=0.4]{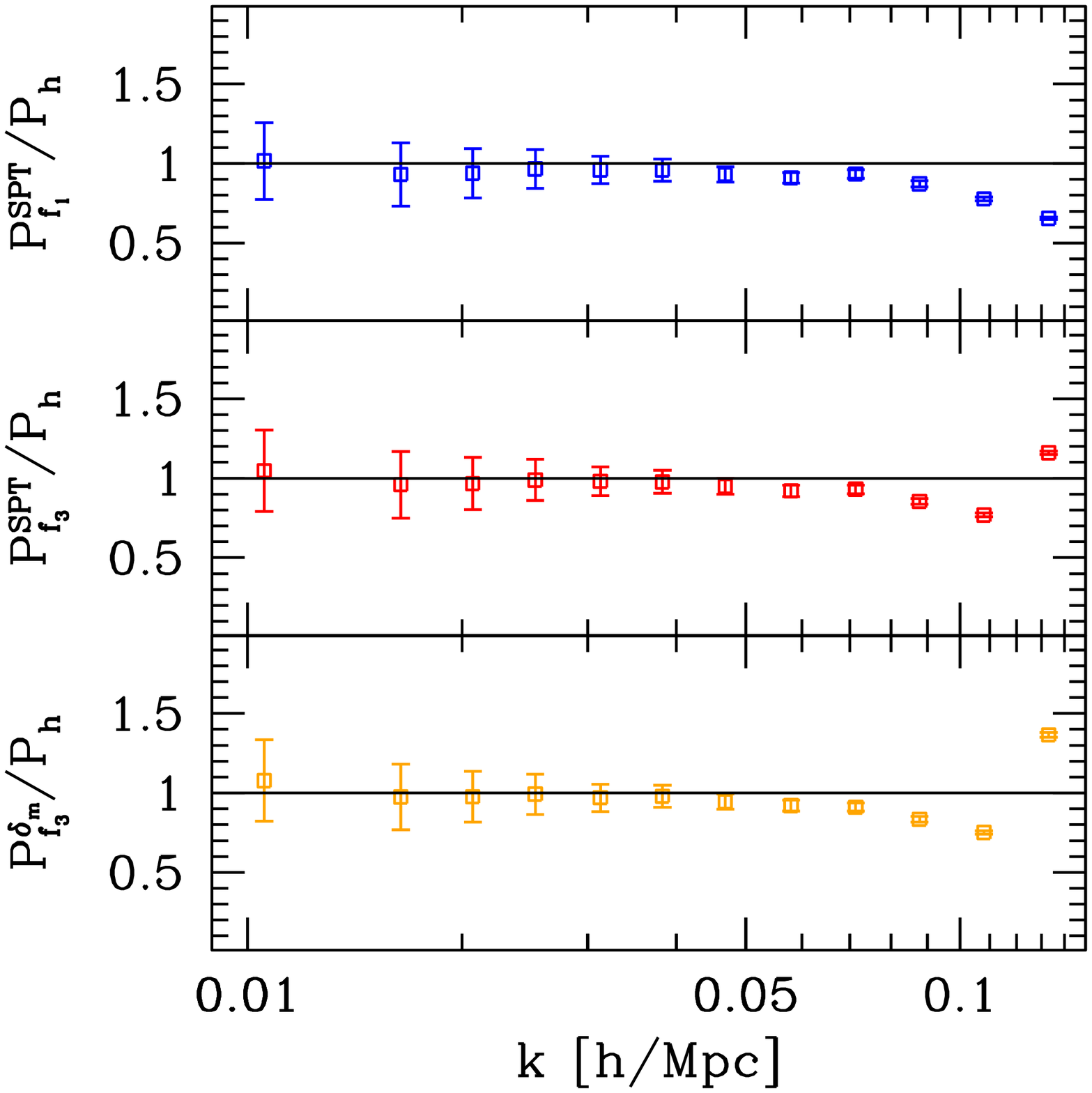}
\includegraphics[scale=0.4]{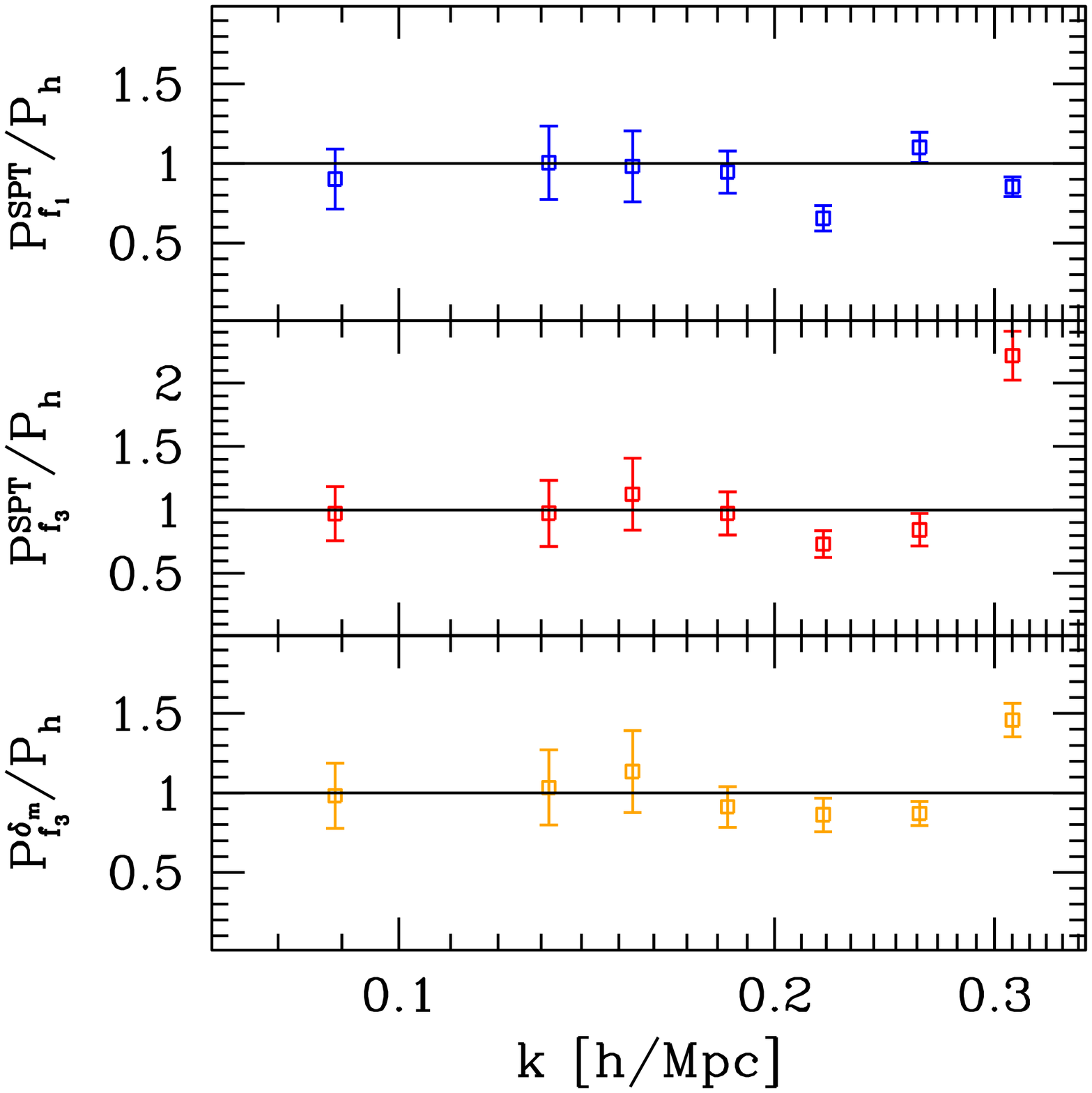}
 \caption{Ratios of \textit{halo power spectra} from our fits to the simulation. Left panel: big simulation volume (mass bin IV, $R=28\ \mathrm{Mpc}/h$), right panel: small simulation volume (mass bin I, $R=12\ \mathrm{Mpc}/h$). From top to bottom: linear fit to the linear matter density; third-order fit to full SPT density; third-order fit to the non-linear matter density field.}
\label{fig:ratio_comp_f11_f31_f31spt}
\end{figure*}
\begin{figure*}
\centering
\includegraphics[scale=0.39]{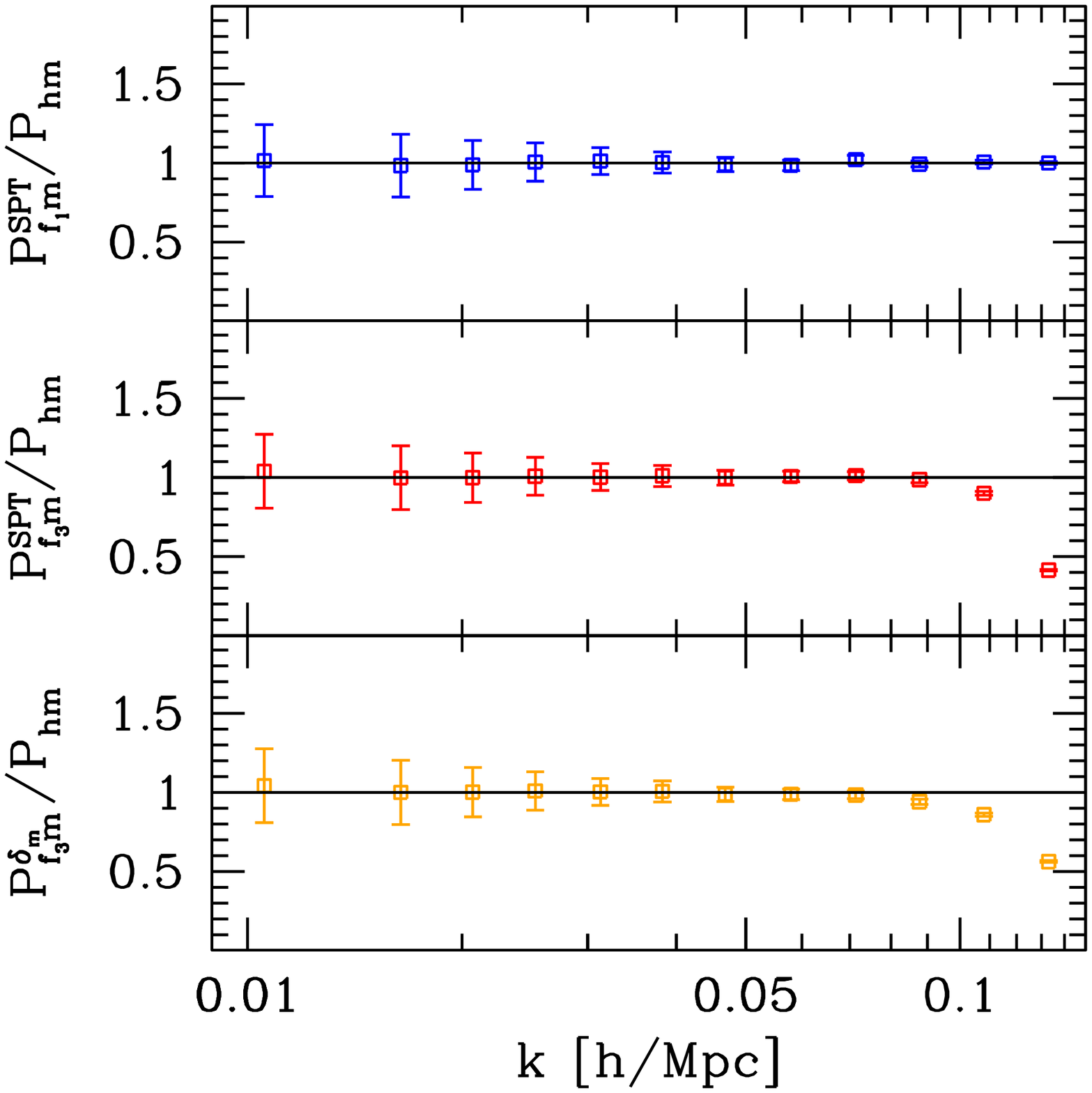}
\includegraphics[scale=0.39]{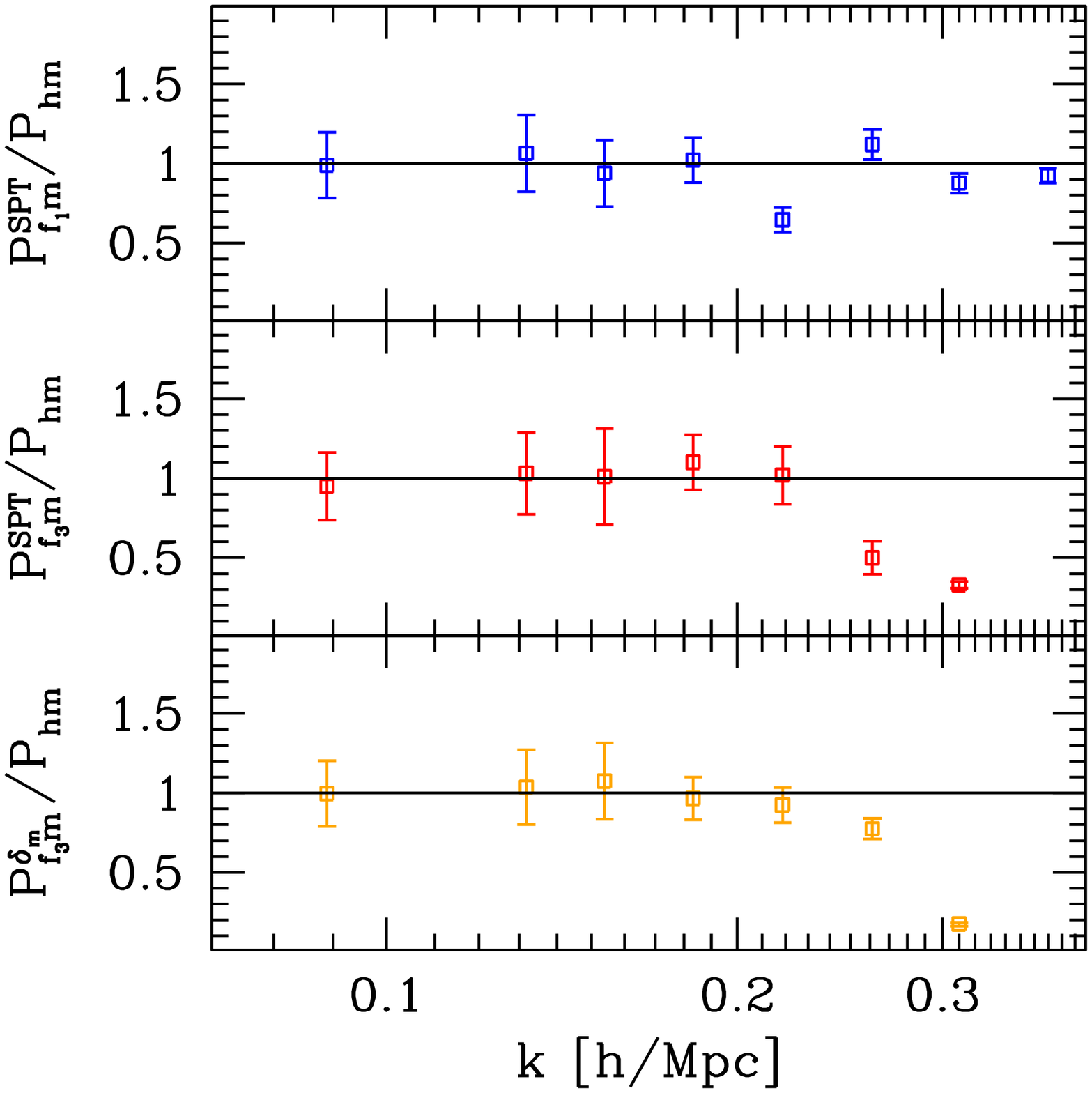}
 \caption{Ratios of \textit{halo-matter cross spectra} from our fits to the simulation. Left panel: big simulation volume (mass bin IV, $R=28\ \mathrm{Mpc}/h$), right panel: small simulation volume (mass bin I, $R=12\ \mathrm{Mpc}/h$). From top to bottom: linear fit to the linear matter density; third-order fit to full SPT density; third-order fit to the non-linear matter density field.}
\label{fig:cross_comp_smbb}
\end{figure*}
\begin{figure*}
\centering
\includegraphics[scale=0.39]{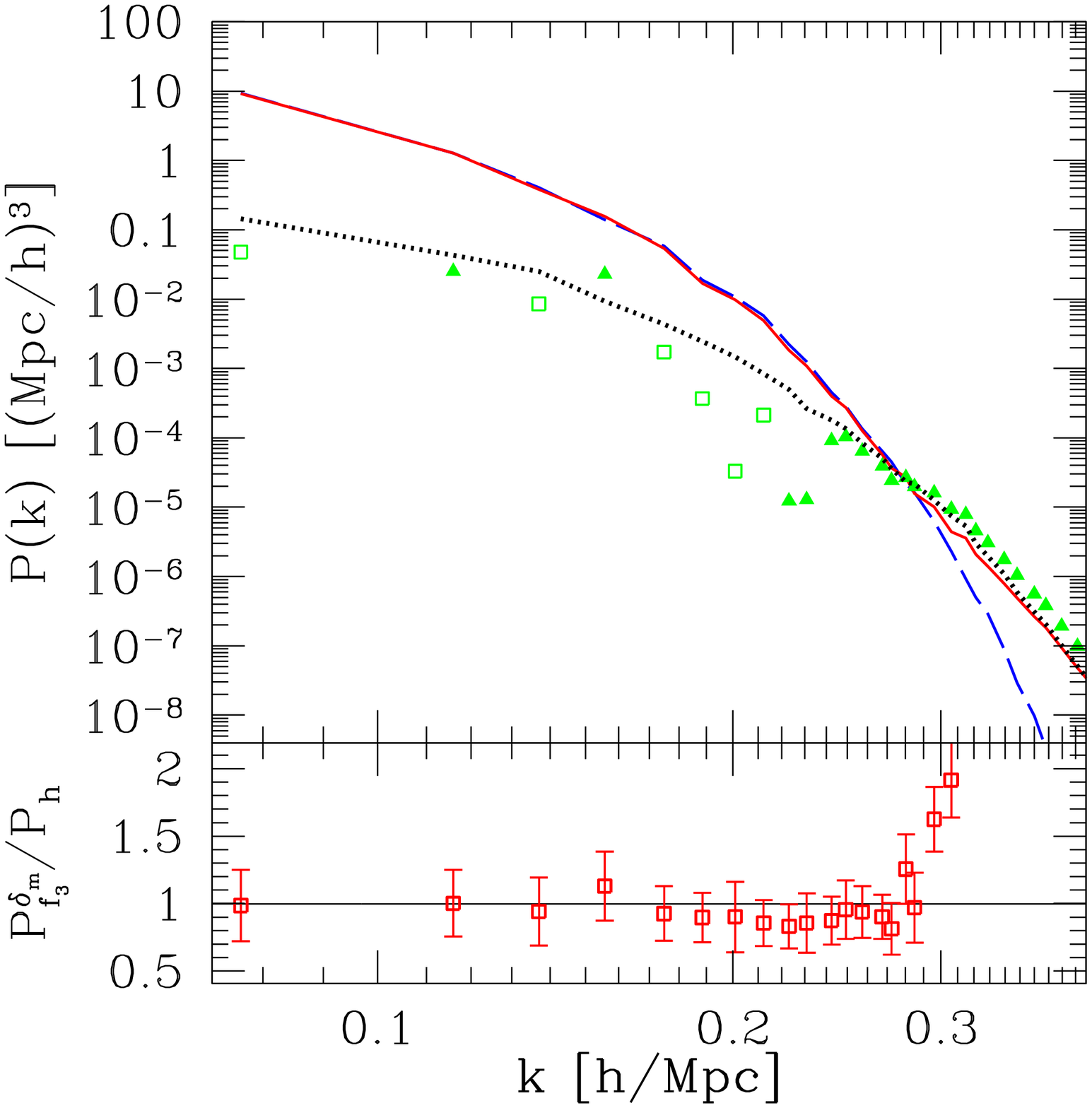}
\includegraphics[scale=0.39]{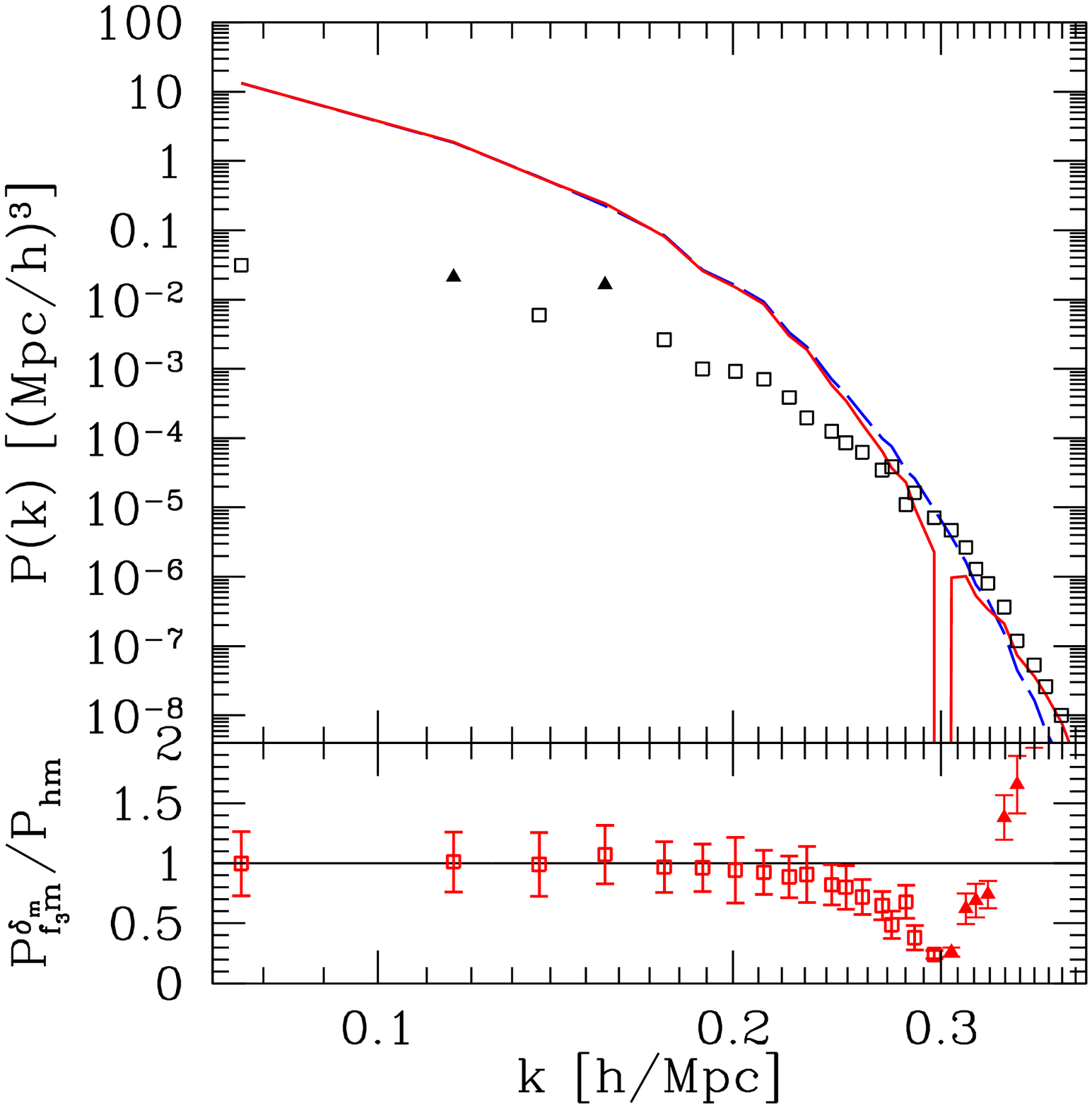}
 \caption{Left: Halo power spectrum and residuals for $\delta_{\mathrm{h,f}_3}(\mathbf{x})$ (upper panel, mass bin I, $R=12\ \mathrm{Mpc}/h$). Dashed blue line: $P_{\mathrm{h}}(k)$, solid red line: $P_{\mathrm{h,f}_3}(k)$, dotted black line: $P_{\Delta \Delta}(k)$, points: the cross-term $ 2\, P_{\Delta\, \delta_{\mathrm{h,f}_3}}(k)$. Open squares show positive values, solid triangles show negative values of the cross-term. Ratio between $P_{\mathrm{h,f}_3}(k) $ and $P_{\mathrm{h}}(k)$ (lower panel). Right: Same for the halo matter cross spectrum. Dashed blue line: $P_{\mathrm{hm}}(k)$, points: $P_{\Delta\, \delta_{\mathrm{m}}}(k)$, solid red line: $P_{\delta_{\mathrm{h},\mathrm{f}_3} \delta_{\mathrm{m}}}(k)$. Note the change in sign at $k \approx 0.3\ h/\mathrm{Mpc}$.}
\label{fig:cross}
\end{figure*}
In this section we test the accuracy of SPT and ELB for the halo power spectrum. In order to calculate the SPT halo power spectra, we have to re-arrange $\delta_{\mathrm{h},f_i}^{\mathrm{SPT}}(\mathbf{x})$ into terms of the same order in $\delta_{\mathrm{1}}^{\mathrm{s}}(\mathbf{x})$, e.g.
\begin{equation}
\delta_{\mathrm{h}}^{\mathrm{(2)}}(\mathbf{k})\equiv b_1 \delta_2^{\mathrm{s}}(\mathbf{x}) + \frac{b_2}{2} (\delta_1^{\mathrm{s}}(\mathbf{x}))^2.
\label{dh2}
\end{equation}
The biased, third-order SPT halo power spectrum then consists of three terms: 
\begin{equation}
P^{\mathrm{SPT}}_{\mathrm{f}_3}(k) \equiv P_{\mathrm{h},11}(k)+ P_{\mathrm{h},22}(k)+2\, P_{\mathrm{h},13}(k),
\end{equation}
with $P_{\mathrm{h},mn}(k)\propto \langle \tilde{\delta}_{\mathrm{h}}^{(m)}(\mathbf{k})\, \tilde{\delta}_{\mathrm{h}}^{(n)}(\mathbf{k'}) \rangle$ (as in Eq. \ref{pgen}). In this way we can make sure that also the halo power spectrum does not contain terms of order higher than $\delta_1^4$.

Fig. \ref{fig:ratio_comp_f11_f31_f31spt} shows the ratio of the reconstructed halo power spectrum to the one from the simulation (left panel: large box, right panel: small box), which has been corrected for (Poissonian) shot noise in the following way:
\begin{equation}
P_{\mathrm{h}}(k)\equiv P^{\mathrm{sim}}_{\mathrm{h}}(k) - \frac{L^3}{\overline{N}} ,
\end{equation}
where $\overline{N}$ is the number of halos in each bin and $L$ is the box size. The top panel on each side corresponds to the power spectrum of $\delta_{\mathrm{h},\mathrm{f}_1}^{\mathrm{SPT}}$, while the middle panel corresponds to $\delta_{\mathrm{h},\mathrm{f}_3}^{\mathrm{SPT}}$. In the lowest panel we consider $\delta_{\mathrm{h},\mathrm{f}_3}$. Two effects can be noticed: In general, both linear and third-order SPT underestimates the power apart from the largest scales in both boxes. However, when $kR \approx \pi$, the unsmoothing effect sets in, eventually leading to a diverging ratio. The lowest panel shows that this behaviour is not caused by SPT, because $\delta_{\mathrm{h},\mathrm{f}_3}^{\mathrm{SPT}}$ and $\delta_{\mathrm{h},\mathrm{f}_3}$ have very similar power spectra. 

We also tried to reverse the order of smoothing and fitting in the following way: Determining the bias parameters from a scatter plot of the unsmoothed\footnote{That is without applying Eq. (\ref{smooth}). Of course, the fields are always smoothed at least on the scale of the grid used in the CIC algorithm.} halo and matter densities, then multiplying the (unsmoothed) matter density with these new bias parameters and applying Eq. (\ref{smooth}) to the resulting biased halo field. The halo power spectra from this fit do not show the unsmoothing effect by construction, but are also incompatible with the simulation halo power spectra on any scale. In fact, not even the linear bias parameter from this new method agrees with the previous estimates, so that the halo power spectrum is different from the simulation even on large scales where the smoothing should have no effect (because $P_{\mathrm{h}} \approx b_1^2 P_{\mathrm{m}}$). We therefore conclude that (not surprisingly) the order of smoothing and fitting can not be reversed, and one has to live with the limitations of the non-linear local bias model that re-introduces some small-scale fluctuations.

We show the ratio of the halo-matter cross spectra in Fig. \ref{fig:cross_comp_smbb}, the order of the panels is the same as in the previous figure. The cross spectrum is not affected by shot noise, and ELB spectra agree with the simulation over a larger range of scales for the big box. Contrary to the halo power spectra, the unsmoothing effect leads to an additional loss in power for the cross spectra on the very small scales, particularly evident for the small box. Note that the best agreement between simulation and fits is actually achieved using the linear bias parameter and the linear matter power spectrum (top panel). For the large box, this is not suprising because the scales in consideration are large enough for linear theory to be a good approximation. For the small box, linear theory is underestimating the power for the small scales, as seen before in Figs. \ref{fig:psum} and \ref{fig:ratio_comp_f11_f31_f31spt}, but only on scales where the cross spectra from the higher-order fits are also too low. This means that the unsmoothing effect (which does not affect the top panel) has an even stronger influence on the halo power spectra from SPT+ELB than the deviation from linear theory on small scales, at least within the accuracy of our simulations.

In order to investigate the systematic effects seen in Figs. \ref{fig:ratio_comp_f11_f31_f31spt} and \ref{fig:cross_comp_smbb}, we now consider the influence of the fit residuals on auto and cross spectra, which include both the intrinsic scatter as well as the unsmoothing effect from the local bias scheme. We define the residual $\Delta(\mathbf{x})$ for the third-order fit by
\begin{equation}
 \Delta(\mathbf{x}) \equiv \delta_{\mathrm{h}}^{\mathrm{s}}(\mathbf{x})-\delta_{\mathrm{h},\mathrm{f}_3}(\mathbf{x}),
\end{equation}
from which 
\begin{align}
P_{\mathrm{h}}(k) &= P_{\mathrm{h,f}_3}(k) + 2\, P_{\Delta\, \delta_{\mathrm{h,f}_3}}(k) + P_{\Delta \Delta}(k), \nonumber \\
P_{\mathrm{hm}}(k) &= P_{\delta_{\mathrm{h},\mathrm{f}_3} \delta_{\mathrm{m}}}(k) + P_{\Delta\, \delta_{\mathrm{m}}}(k),
\end{align}
follows for the halo power spectrum and the halo matter cross spectrum. The left panel of Fig. \ref{fig:cross} shows the different terms for $P_{\mathrm{h}}(k)$: $P_{\mathrm{h,f}_3}(k)$ (solid red) is very close to $P_{\mathrm{h}}(k)$ (dashed blue) until $k \approx 0.3\ h/\mathrm{Mpc}$. The cross-term $2\, P_{\Delta\, \delta_{\mathrm{h,f}_3}}(k)$ (green points) can have both positive and negative values. This is indicated in the figure by the different symbols: open squares for positive values, solid triangles for negative values. Finally, $P_{\Delta \Delta}(k)$ is shown with a black dotted line. Generally, $2\, P_{\Delta\, \delta_{\mathrm{h,f}_i}}(k)$ and $P_{\Delta \Delta}(k)$ have a small amplitude, but their contribution becomes important on smaller scales. The lower panel shows the ratio of the fit and the simulated halo power spectrum for comparison. The region where $P_{\mathrm{h,f}_3}(k)$ is systematically lower than $P_{\mathrm{h}}(k)$ roughly corresponds to the region where the cross-term is positive.

The right panel of Fig. \ref{fig:cross} shows the terms contributing to $P_{\mathrm{hm}}(k)$ (blue dashed): the cross spectrum of the fit and the dark matter (red solid) and the cross spectrum of the residuals and the dark matter (black points, squares positive values, triangles negative values). In contrast to the fitted halo power spectrum which rises on small scales, the fitted cross spectrum becomes negative at around the same scales, indicated by the ``hole" in the red solid line (for smaller scales we then show $-P_ {\delta_{\mathrm{h},\mathrm{f}_3}\delta_{\mathrm{m}}}$). The lower panel shows the ratio of the fit and the simulation cross spectrum. As before, squares correspond to positive values of this ratio, and triangles show where it is negative. 

We conclude that the halo auto and cross spectra computed with the ELB deviate from the simulation on scales where $kR \approx \pi$. This affects especially the auto spectra, and is not related to using SPT instead of the non-linear matter field. 
\subsubsection{Biased SPT on Large Scales}
\label{ssec:beff}
\begin{table*}
\centering
\caption{Comparing $b_1$, $\bar{b}_1$, $b_{\mathrm{eff}}$, $b^{\mathrm{H}}_{\mathrm{eff}}$ and $b^{\mathrm{spt}}_{\mathrm{eff}}$ for different mass bins and smoothing scales $R$. $\langle \cdot \rangle$ here denotes the average over a specific $k$-range (see text). $R$ is given in Mpc$/h$. Errors are jackknife errors using 8 subsamples.}
\begin{tabular}{c|c|c|c|c|c|c}
\hline Mass bin &$R$ & $b_1$& $\bar{b}_1$& $\langle b_{\mathrm{eff}} \rangle$ &$\langle b^{\mathrm{H}}_{\mathrm{eff}} \rangle$ & $\langle b^{\mathrm{spt}}_{\mathrm{eff}} \rangle$\\
\hline

I   & 6  & 0.773 $\pm$ 0.005 & 0.573 $\pm$ 0.011 & 0.684 $\pm$ 0.051 & 0.606 $\pm$ 0.040 & 0.623 $\pm$ 0.028\\  
II  & 6  & 0.928 $\pm$ 0.011 & 0.680 $\pm$ 0.013 & 0.756 $\pm$ 0.057 & 0.484 $\pm$ 0.067 & 0.800 $\pm$ 0.032\\ 
III & 6  & 1.476 $\pm$ 0.030 & 1.108 $\pm$ 0.043 & 1.198 $\pm$ 0.091 & 0.722 $\pm$ 0.112 & 1.100 $\pm$ 0.061\\ 
I   & 12 & 0.719 $\pm$ 0.009 & 0.694 $\pm$ 0.001 & 0.684 $\pm$ 0.051 & 0.545 $\pm$ 0.031 & 0.703 $\pm$ 0.044\\ 
IV  & 28 & 1.279 $\pm$ 0.008 & 1.278 $\pm$ 0.001 & 1.253 $\pm$ 0.076 & 1.056 $\pm$ 0.042 & 1.243 $\pm$ 0.059\\ 
IV  & 50 & 1.244 $\pm$ 0.028 & 1.241 $\pm$ 0.028 & 1.253 $\pm$ 0.076 & 1.185 $\pm$ 0.060 & 1.237 $\pm$ 0.054\\ 
V   & 28 & 2.040 $\pm$ 0.023 & 2.037 $\pm$ 0.022 & 1.958 $\pm$ 0.119 & 1.553 $\pm$ 0.295 & 1.947 $\pm$ 0.095\\ 
V   & 50 & 2.029 $\pm$ 0.042 & 2.016 $\pm$ 0.039 & 1.958 $\pm$ 0.119 & 1.697 $\pm$ 0.111 & 1.972 $\pm$ 0.089\\ 
VI  & 28 & 3.704 $\pm$ 0.054 & 3.615 $\pm$ 0.049 & 3.520 $\pm$ 0.222 & 4.739 $\pm$ 0.134 & 3.662 $\pm$ 0.159\\
VI  & 50 & 3.761 $\pm$ 0.088 & 3.739 $\pm$ 0.087 & 3.520 $\pm$ 0.222 & 4.461 $\pm$ 0.296 & 3.877 $\pm$ 0.155\\
\hline
\end{tabular}
\label{tab:bcomp}
\end{table*}
\citet{Heavens98} discuss two effects on the halo power spectrum that are caused by using SPT. The basis for their analysis is as follows: as for the case of the one-loop matter power spectra, one can also express the SPT halo power spectra as integrals over products of the linear matter power spectrum and the bias parameters \citep{Jain:1993jh}. The $P_{\mathrm{h},22}$ and $P_{\mathrm{h},13}$ terms defined before can then be written as
\begin{align}
P_{\mathrm{h},22}(k) &=\ 2 \int  \frac{d^3q}{(2\pi)^3} P_{11}(q) P_{11}(|\mathbf{k-q}|)\ \times \nonumber \\ 
& \qquad \quad \left[  b_1 F_2^{(s)}(\mathbf{q},\mathbf{k-q}) + \frac{b_2}{2} \right ]^2, 
 \nonumber \\ 
P_{\mathrm{h},13}(k)&= 6\, b_1 P_{11}(k) \int \frac{d^3q}{(2\pi)^3} P_{11}(q)\left[ b_1 F_3^{(s)}(\mathbf{q},\mathbf{-q},\mathbf{k}) \right. \nonumber \\
& \qquad \qquad \ \quad \  \ \left. + \frac{b_3}{6} + b_2 F_2^{(s)}(\mathbf{-q},\mathbf{k}) \right],
\label{phtheo}
\end{align}
and $P_{\mathrm{h},11}(k)=b_1^2 P_{11}(k)$.
By taking the limit $k\rightarrow 0$ of the foregoing equations, one can study the behaviour of the SPT halo power spectrum on large scales:
\begin{enumerate}
\item \textbf{Is the large-scale bias not $b_1$?} While the linear bias model should be valid on large scales, \cite{Heavens98} predict that the large-scale bias is not $b_1$, but can be approximated by an effective bias
\begin{equation}
b^{\mathrm{H}}_{\mathrm{eff}}= \sqrt{b_1^2 + b_1 \left(\frac{68}{21}\, b_2 +b_3 \right) \sigma^2_R},
 \label{bheav}
\end{equation}
where $\sigma^2_R$ is the variance of the smoothed linear density field at redshift 0:
\begin{equation}
 \sigma^2_R \equiv \int \frac{\mathrm{d}^3q}{(2 \pi)^3} P_{11}(q)\, e^{-(qR)^2}.
\end{equation}
Eq. (\ref{bheav}) tells us that the difference between $b^{\mathrm{H}}_{\mathrm{eff}}$ and $b_1$ depends on the choice of the smoothing scale through $\sigma^2_R$. However, $b_1$ is supposed to describe the large-scale behaviour of the power spectrum, where the smoothing scale should not have any effect. The value of $\sigma^2_R$ will be large for smaller $R$ and $b_2$ is negative in many cases, so the term in parenthesis can even become negative, leading to $b^2_{\mathrm{eff}}<0$. (Although this never happens for the smoothing scales we investigate.) We can compare this with the effective bias directly from the simulation power spectra
\begin{equation}
b_{\mathrm{eff}} \equiv \sqrt{\frac{P_{\mathrm{h}}}{P_{\mathrm{m}}}}
\label{beff_rat}
\end{equation}
on large scales ($k\leq 0.02\ h/\mathrm{Mpc}$ for the large box and $k\leq 0.15\ h/\mathrm{Mpc}$ for the small box), where this ratio is found to be constant within the errorbars. Note that $P_{\mathrm{h}}$ has been corrected for shot noise, and for this specific ratio, the dependence on the smoothing scale cancels out. Examples of the different bias parameters are shown in Table \ref{tab:bcomp}, for mass bins in both boxes and several smoothing scales. Here, $b_1$ and $\bar{b}_1$ are the linear bias parameters from the third-order fits, and $b_{\mathrm{eff}}^{\mathrm{H}}$ was also calculated from these values. We always find the large-scale bias to be very close to $b_1$ for reasonable\footnote{Meaning that the SPT assumptions about the smallness of $\delta$ are still valid at $z=0$, requiring $R \geq 8\, \mathrm{Mpc}/h$.} smoothing scales. Note that the value for $b_{\mathrm{eff}}$ for the small box (mass bins I-III) is only an approximation for the large-scale bias, because the box is so small. Comparing $b_1$, $\bar{b}_1$ and $\langle b_{\mathrm{eff}} \rangle$ for the low mass bins, it is clear that the different estimates do not agree with each other if the smoothing scale gets too small ($R=6\ \mathrm{Mpc}/h$). Note also that exchanging $P_{\mathrm{h}}$ with one of the fitted halo power spectra in Eq. (\ref{beff_rat}) does not significantly affect the value of $b_{\mathrm{eff}}$ (exchanging $P_{\mathrm{m}} \rightarrow P_{\mathrm{SPT}}$ for the SPT fit as well). 

\begin{figure*}
\centering
\includegraphics[scale=0.35]{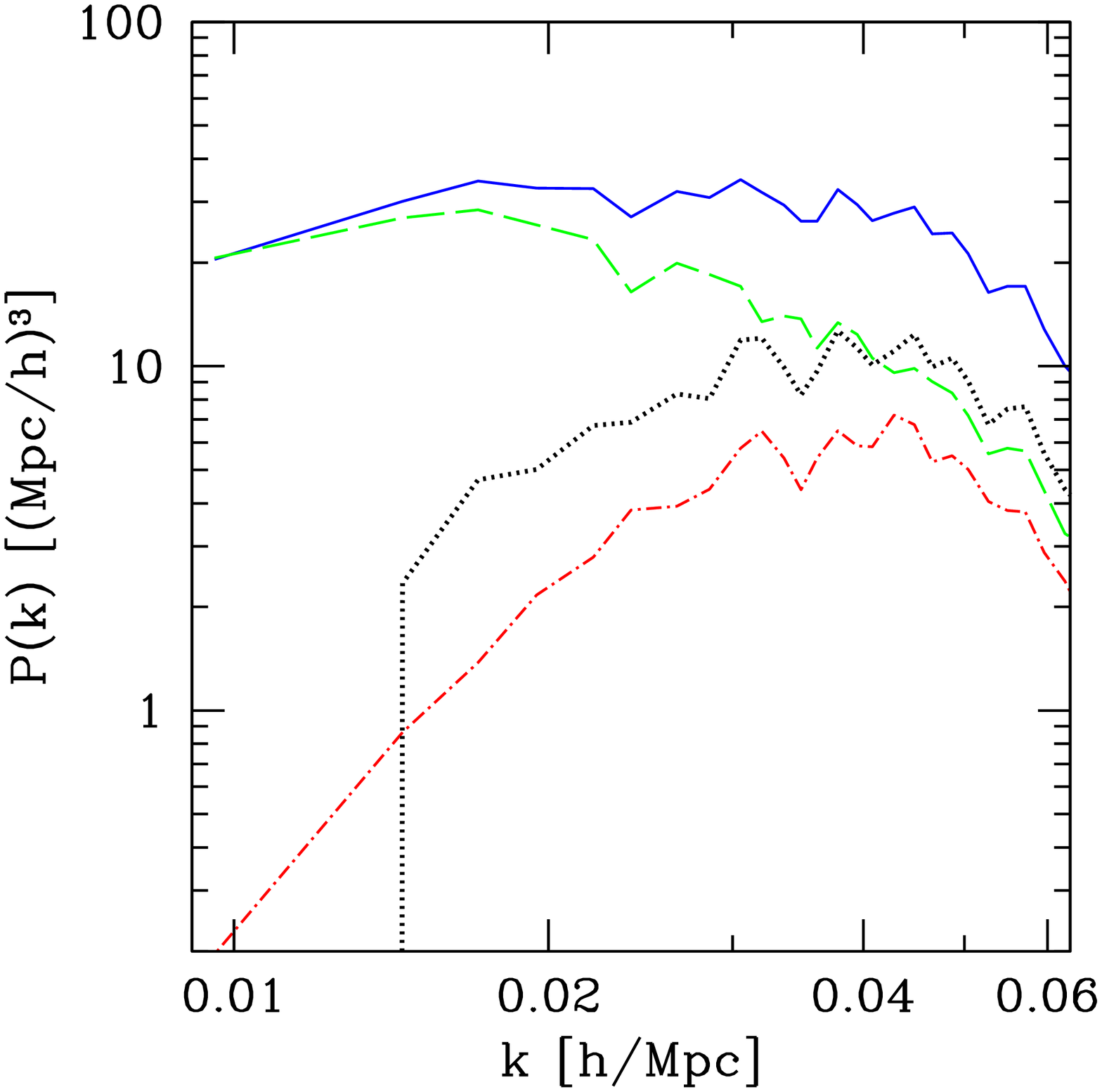}
\includegraphics[scale=0.35]{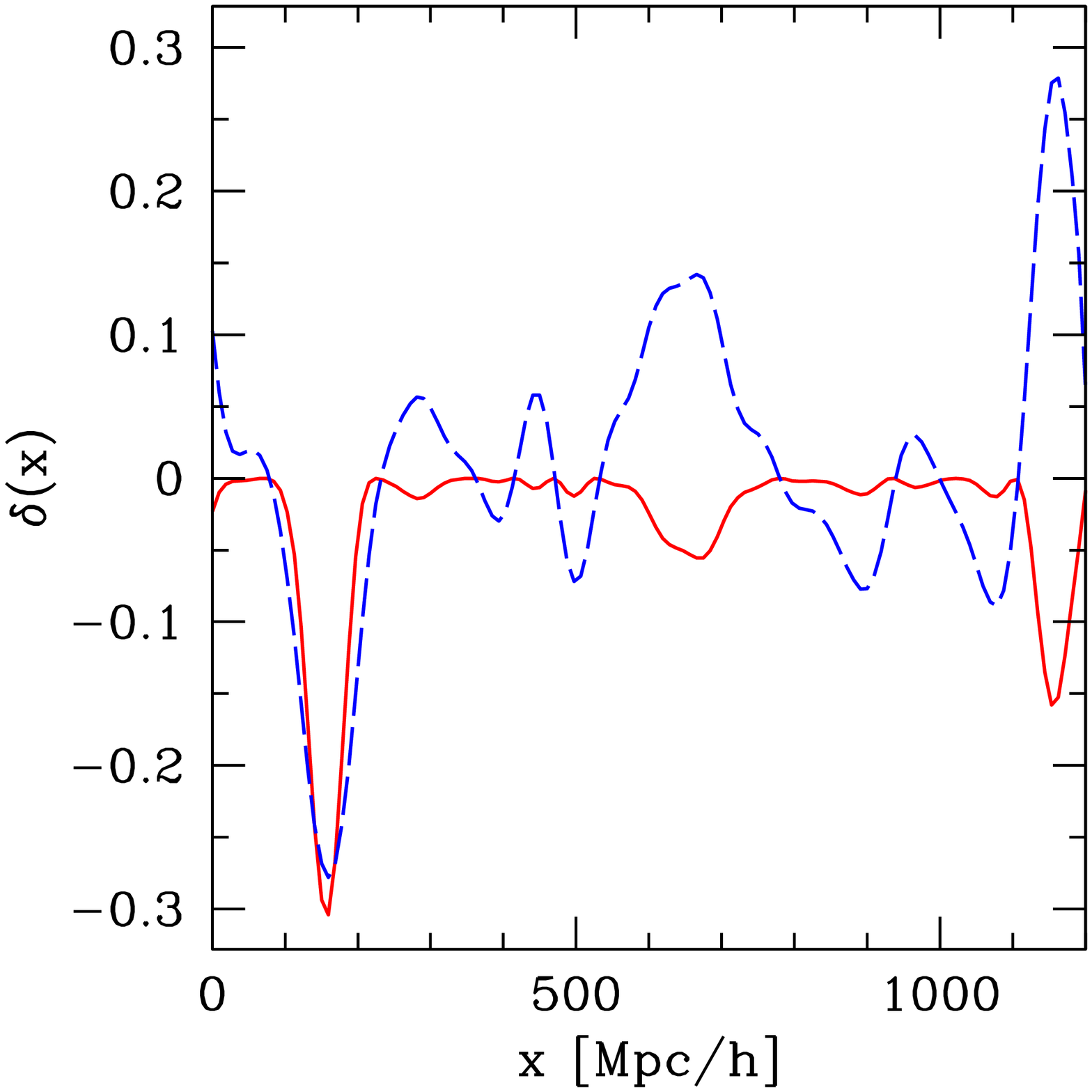}
\caption{Left: Different contributions to $P_{\mathrm{h},22}(k)$ (see text) and their sum (blue solid) for mass bin IV with $R=28\ \mathrm{Mpc}/h$. The constant power on large scales is dominated by $\frac{b_2^2}{4} \langle \tilde{\delta}_1^2(\mathbf{k}) \tilde{\delta}_1^2(\mathbf{k'})\rangle$ (green line). Right: The contribution $A \cdot \delta_1^2(\mathbf{x})$ (solid red, with $A=\frac{b_2}{2}\cdot 5$ for enhanced visibility), which coincides with the peaks and troughs of the underlying SPT matter density field (dashed blue) as suggested by Heavens et al. (1998).}
\label{fig:peaks}
\end{figure*}   
Extending the comparison to $b_{\mathrm{eff}}^{\mathrm{H}}$ reveals a large discrepancy to the previous values over the whole mass range and different smoothing scales. Even though the errors on $b_{\mathrm{eff}}^{\mathrm{H}}$ are rather large (owing to the large errors in $\bar{b}_3$), the estimates are quite different from the fit parameters and $b_{\mathrm{eff}}$.

In concordance with the derivation of Eq. (\ref{bheav}) in \citet{Heavens98}, we define an additional effective bias for the SPT fit:
\begin{equation}
b^{\mathrm{spt}}_{\mathrm{eff}} \equiv \sqrt{\frac {P_{\mathrm{f}_3}^{\mathrm{SPT}} } { P_{11} } } ,
\end{equation}
which is also averaged over the same scales as $b_{\mathrm{eff}}$. The difference to Eq. (\ref{beff_rat}) comes because the denominator contains only the linear power spectrum. However, the values of $b^{\mathrm{spt}}_{\mathrm{eff}}$ do not differ much from $b_{\mathrm{eff}}$ because on large scales $P_{\mathrm{m}} \approx P_{\mathrm{SPT}} \approx P_{11}$.

We conclude that $b_{\mathrm{eff}}$ matches the fit parameters $b_1$ and $\bar{b}_1$ within the errorbars. Eq. (\ref{bheav}) was derived neglecting the contribution of $P_{\mathrm{h},22}(k)$, which seems not to be a valid approximation, as the values for $b_{\mathrm{eff}}^{\mathrm{H}}$ do not agree with $b_{\mathrm{eff}}$ and $b_{\mathrm{eff}}^{\mathrm{spt}}$. It is also important to note that the agreement of the different bias estimations gets worse when smaller smoothing scales are considered. This implies that using the bias parameters to infer halo masses (by choosing $R$ to correspond to the Lagrangian radius of the halo) can lead to wrong results.
\item \textbf{Is the SPT halo power spectrum constant on very large scales?} From Eq. (\ref{phtheo}) we see that the $P_{\mathrm{h},22}(k)$-term is not directly proportional to $P_{11}(k)$ which falls off as $k\rightarrow 0$. It follows that this term will eventually dominate, and it can be shown to lead to a constant halo power spectrum on very large scales. We can study this behaviour using the SPT halo density contrast from the third-order fit. Even with our large simulation volume, we can only see that $P_{\mathrm{h},22}(k)$ tends to a constant, but the scales where it actually dominates are out of reach. However, we can look in more detail at the $P_{\mathrm{h},22}(k)$-term, to find out why it becomes constant. From Eq. (\ref{dh2}), we can see that there a three terms that contribute, which are shown in the left panel of Fig. \ref{fig:peaks}: $b_1^2 P_{22}(k)$ (red dot-dashed), the cross term $b_1 b_2 \langle \tilde{\delta}_2^{\mathrm{s}}(k) \tilde{\delta}_1^2 (k')\rangle$ (black dotted) and $\frac{b_2^2}{4} \langle \tilde{\delta}_1^2(k) \tilde{\delta}_1^2(k') \rangle$ (green dashed). The latter term clearly dominates $P_{\mathrm{h},22}(k)$ (blue solid line) on large scales. \cite{Heavens98} suggest that this constant power comes from the peaks and troughs of the underlying density field, which we can confirm: The right panel of Fig. \ref{fig:peaks} shows the SPT matter density contrast $\delta_{\mathrm{SPT}}^{\mathrm{s}}(\mathbf{x})$ and $\frac{b_2}{2}\, \delta_1^2({\mathbf{x}})$ along a line in the simulation volume. The latter term can have either sign depending on $b_2$, here we show the mass bin IV where $b_2<0$. Note that $\frac{b_2}{2}\, \delta_1^2$ has been multiplied by a factor of 5 to make the effect more visible.
\end{enumerate}
\section{Conclusion}
\label{sec:conc}
In this work we have followed a novel approach by evaluating the SPT expansion of the mass density and velocity fields up to third order, starting from the same realisations of the linear density field that has been used to generate the initial conditions of two N-body simulations. This allowed us to make a point-by-point comparison between the non-linear mass and halo overdensities (using the Eulerian local bias model), while past studies have only focussed on two- or three-point statistics. Our results can be summarized as follows:
\begin{itemize}
\item We found that SPT provides a good approximation to the density field up to redshift 0, for smoothing scales $R \geq 8 \ \mathrm{Mpc}/h$. This ensures that the linear density contrast is typically less than one.
\item We have compared the redshift evolution of the non-linear matter density field of the simulations and SPT with the prediction of linear theory and the spherical collapse model. We found that SPT is very close to the simulated density for all redshifts, while both linear theory and the spherical collapse model are in poor agreement with the simulations. The lognormal model by \citet{2001ApJ...561...22K} gives a good PDF for the non-linear density contrast but fails when used for a point-by-point comparison.
\item In Eq. (\ref{fit_nl_pol}), we presented a simple fitting formula for the non-linear density contrast as a function of the linear density contrast, which is accurate at the per cent level over the full range of redshifts available to us ($0\leq z\leq 10$).
\item We compared the SPT matter power spectra with the linear and non-linear matter power spectra from the simulation. On large scales, linear theory provides a good approximation to the simulated matter power spectra, but SPT is superior on smaller scales, up to the maximum wavenumber we can probe, $k \approx 0.3\ h/\mathrm{Mpc}$ at redshift 0.
\item Assuming a deterministic, Eulerian local bias model (ELB) with up to 3 free parameters, we obtained values for these parameters by fitting polynomials to a scatter plot of the smoothed matter and halo density contrast. We find that the third-order bias model is always preferred by the data over models with less parameters. The reconstructed halo density is similar to the simulation, but the ELB can not accurately reproduce the simulated field. 
\item We found that the bias parameters from fitting the halo distribution to the simulated matter density field and to the SPT density field generally differ, but the corresponding power spectra are very similar. The mass dependence of the bias parameters shows the same trend as theoretical predictions based on the peak-background-split approach.
\item We compared both the halo-halo power spectrum and the halo-matter cross spectrum from the fits with the ones from the simulation. The cross spectrum is in better agreement with the simulation than the auto spectrum which deviates from the simulation on scales much larger than the smoothing radius. This is not related to using SPT instead of the non-linear matter field. 
\item We have investigated two effects on the halo power spectra, which have been predicted by analytic considerations of SPT and the local bias model. First, we estimated the large-scale bias $b_{\mathrm{eff}}$ using the halo and mass power spectra from both the simulation and the fits. This large-scale bias is compatible with the linear bias parameter obtained from the polynomial fit if we make sure that the SPT assumptions are not violated, choosing the smoothing radius $R$ such that $\delta \ll 1$. This suggests that the effective bias does not require perturbative corrections, contrary to previous results based on SPT. Second, we determined the origin of the constant shot-noise term on very large scales, which is caused by $\frac{b_2^2}{4}\langle \tilde{\delta}_1^2 \tilde{\delta}_1^2 \rangle$ as predicted in \citet{Heavens98}.
\end{itemize}
In summary, our study shows that SPT is a suitable approximation for the matter field even at redshift 0, provided a large enough smoothing radius is adopted. However, the Eulerian local bias model can not fully describe the halo density field, which is most evident from our point-by-point comparison in Fig. \ref{fig:dhvsfdmdm_lowdens}.  
\section{Acknowledgements}
We acknowledge support through the SFB-Transregio 33 "The Dark Universe" by the Deutsche Forschungsgemeinschaft (DFG).
\nocite{2007AAS...211.9108J} 
\nocite{2009ApJ...691..569J} 

 \bibliography{final}
 \bibliographystyle{mn2e}

\end{document}